\documentclass[journal,onecolumn,12pt]{IEEEtran}
\renewcommand{\baselinestretch}{1.5}
\usepackage{cite}
\usepackage{amssymb}
\usepackage{amsmath}
\usepackage{graphicx}
\usepackage{epsfig}
\usepackage{amsthm}
\usepackage{subfigure}

\DeclareMathOperator*{\argmax}{argmax}
\DeclareMathOperator*{\argmin}{argmin}

\begin{document}
\thispagestyle{empty} \vskip 1cm

\begin{figure}
	\centering
	\includegraphics[scale=0.225]{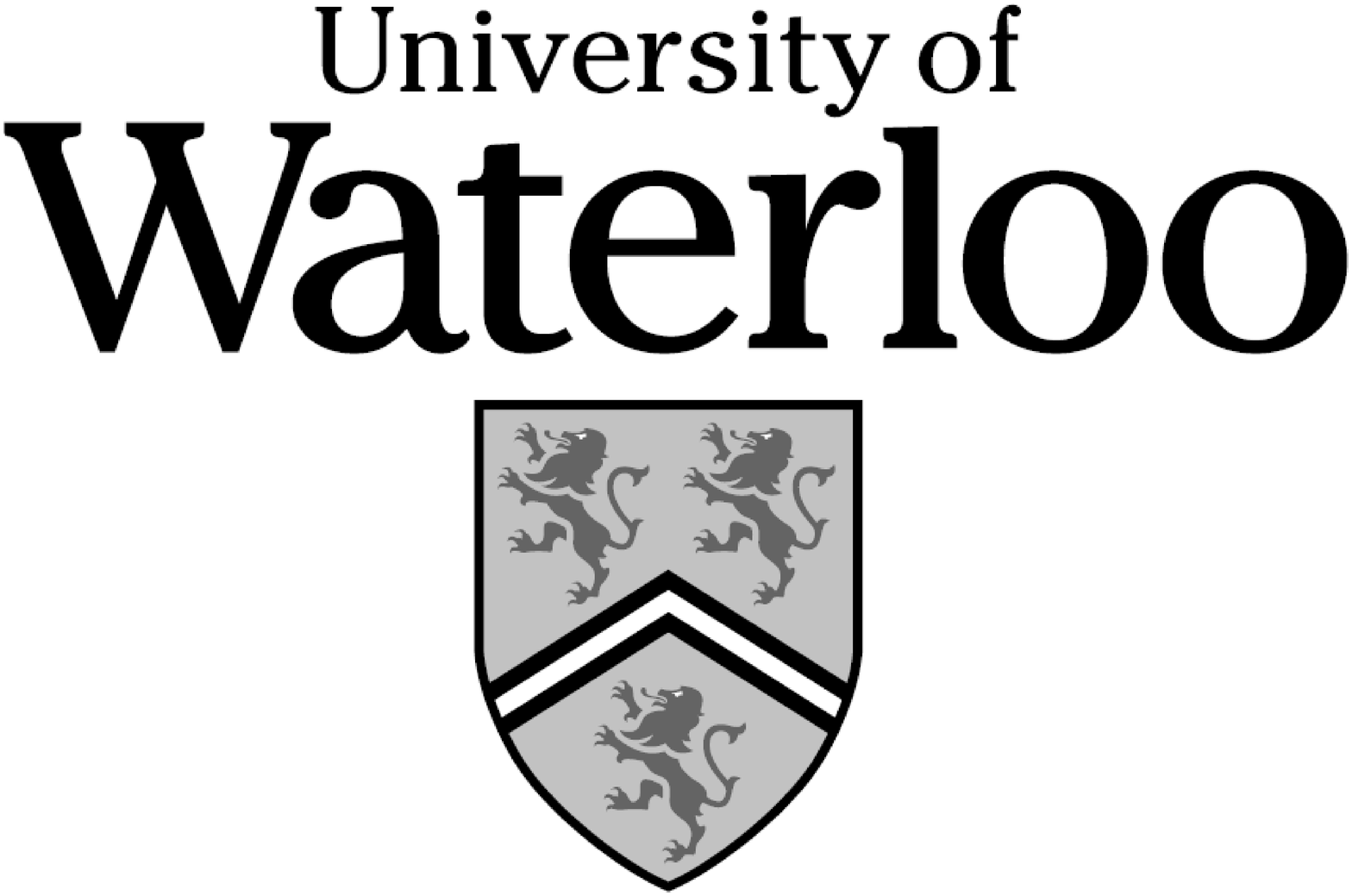} 
\end{figure}


\begin{center}
\vskip 5.5cm \Large \bf Path Diversity over Packet Switched Networks: \\ Performance Analysis and Rate Allocation \\

\vskip 0.5cm \large Shervan Fashandi, Shahab Oveis Gharan and Amir K. Khandani

\vskip 0.5cm

\centering{ {\small Electrical and Computer Engineering Department\\
University of Waterloo, Waterloo, ON, Canada\\
E-mail:\{sfashand,shahab,khandani\}@cst.uwaterloo.ca\\
}}

\vskip 0.5cm

Technical Report UW-E\&CE\#2008-09\\
May 2008

\end{center}

\pagenumbering{arabic} \setcounter{page}{0}
\title{Path Diversity over Packet Switched Networks: \\ Performance Analysis and Rate Allocation}
\author{Shervan Fashandi, Shahab Oveis Gharan and Amir K. Khandani,~\IEEEmembership{Member,~IEEE}}
\maketitle
\begin{abstract}
Path diversity works by setting up multiple parallel connections
between the end points using the topological path redundancy of the
network. In this paper, \textit{Forward Error Correction} (FEC) is
applied across multiple independent paths to enhance the end-to-end
reliability. Network paths are modeled as erasure Gilbert-Elliot
channels~\cite{Bolot1999,Bolot1996,Nguyen2003,Nguyen2004,Leannec1999}. It is known that over any
erasure channel, \textit{Maximum Distance Separable} (MDS) codes
achieve the minimum probability of irrecoverable loss among all
block codes of the same size~\cite{Fashandi2008isit,Fashandi20083}. Based on the adopted model for the error behavior,
we prove that the probability of irrecoverable loss for MDS codes
decays exponentially for an asymptotically large number of paths.
Then, optimal rate allocation problem is solved for the asymptotic case
where the number of paths is large. Moreover, it is shown that in
such asymptotically optimal rate allocation, each path is assigned a
positive rate \textit{iff} its quality is above a certain threshold.
The quality of a path is defined as the percentage of the time it
spends in the bad state. Finally, using dynamic programming, a
heuristic suboptimal algorithm with polynomial runtime is proposed
for rate allocation over a finite number of paths. This algorithm
converges to the asymptotically optimal rate allocation when the
number of paths is large. The simulation results show that the
proposed algorithm approximates the optimal rate allocation (found
by exhaustive search) very closely for practical number of paths,
and provides significant performance improvement compared to the
alternative schemes of rate allocation.\footnote{Financial support
provided by Nortel and the corresponding matching funds by the
Natural Sciences and Engineering Research Council of Canada (NSERC),
and Ontario Centres of Excellence (OCE) are gratefully
acknowledged.}
\end{abstract}


\section{Introduction}
\label{section:Introduction} \IEEEPARstart{I}{n} recent years,
\textit{path diversity} over the Internet has received significant
attention. It has been shown that path diversity has the ability to
simultaneously improve the end-to-end rate and
reliability~\cite{Han20062,Nguyen2003,Mao2005,Fashandi2007Glob}. In
a dense network like the Internet, it is usually possible to find
multiple independent paths between most pairs of nodes~\cite{Han2006,Han2004,Spring2004,Teixeira2003,Barbasi1999,Andersen2001}. 
A set of paths are defined to be independent if their corresponding packet loss and
delay characteristics are independent. Clearly, disjoint paths would be independent too~\cite{Liang2001,Nelakuditi2004,Nguyen2003,Nguyen2004,Han2006,Han20062,Han2004,Andersen2003}. Even when 
the paths are not completely disjoint, their loss and delay patterns may show a high degree of independence 
as long as the nodes and links they share are not congestion points or bottlenecks~\cite{Liang2001,Nelakuditi2004,Han2006,Han2004,Nguyen2003,Andersen2001,Andersen2003,Teixeira2003}.
In this paper, \textit{Forward Error Correction} (FEC) is applied across multiple independent paths. Based on this model, we show that path diversity significantly enhances the performance of FEC.

In order to apply path diversity over any packet switched network, two problems need to be addressed: i) setting up multiple independent paths between the end-nodes, ii) utilizing the given independent paths to improve the end-to-end throughput and/or reliability. In this paper, we focus on the second problem only. However, it should be noted that the first problem has also received significant attention in the literature (see~\cite{Andersen2001,Chun2004,Han2005,Han2006,Han20062,Han2004,Guo2003,Akella2008,Akella2004,Andersen2003,Srinivasan2007,Cha2006}). In case the end-points have enough control over the path selection process, the centralized and distributed algorithms in references~\cite{Eppstein1994} and~\cite{Ogier1993} can be used to find multiple disjoint paths over a large connected graph. However, applying such algorithms over the Internet requires modification of IP routing protocol and extra signaling between the nodes (routers). Of course, modifying the traditional IP network is extremely costly. To avoid such an expense, overlay
networks are introduced~\cite{Andersen2001,Andersen2003,Clarck2006}. The basic idea of overlay networks is to equip very few nodes (smart nodes) with the desired new functionalities while the rest remain unchanged. The smart nodes form a virtual network connected through virtual or logical links on top of the actual network. Thus, overlay nodes can be used as relays to set up independent paths between the end nodes~\cite{Guo2003,Cha2006,Cha2007,Srinivasan2007,Akella2004}. Han et. al have experimentally studied the number of available disjoint paths in the Internet using overlay networks~\cite{Han2006}. They have also discussed the impact of network path diversity on the performance of overlay networks~\cite{Han2004,Han2005}. Reference~\cite{Chun2004} addresses the problem of distributed overlay network
design based on a game theoretical approach. Many other researchers have tried to optimize the design of overlay networks such that they offer the maximum degree of path diversity~\cite{Guo2003,Srinivasan2007,Cha2006,Cha2007}. Moreover, the idea of \textit{multihoming} is proposed to set up extra independent paths between the end-points~\cite{Akella2008,Akella2004}. In this technique, the end users are connected to more than one \textit{Internet Service Providers} (ISP's) simultaneously. It is shown that combining multihoming with overlay assisted routing can improve the end-to-end performance considerably~\cite{Akella2004}. In the cases where the backbone network partially consists of optical links between the nodes, each optical fiber conveys tens of independent channels (tones). There has been efforts to take advantage of this inherent physical layer diversity in optical networks~\cite{Cha2007}.

Recently, path diversity is utilized in many applications
(see~\cite{Karrer2003, Apostolopoulos2002, Nguyen2004,AkamaiSureRoute,Ghanassi2006}).
Reference~\cite{Apostolopoulos2002} combines multiple description
coding and path diversity to improve quality of service (QoS) in
video streaming. Packet scheduling over multiple paths is addressed
in~\cite{Chakareski2003} to optimize the rate-distortion function of
a video stream. Reference~\cite{Ghanassi2006} utilizes path diversity 
to improve the quality of Voice over IP streams. According to~\cite{Ghanassi2006}, sending 
some redundant voice packets through an extra path helps the receiver 
buffer and the scheduler optimize the trade-off between the maximum tolerable 
delay and the packet loss ratio~\cite{Ghanassi2006}. In~\cite{Han20062}, multipath routing of TCP packets
is applied to control the congestion with minimum signaling
overhead. \textit{Content Distribution Networks} (CDN's) can also take advantage of path diversity for 
performance improvement. CDN's are a special type of overlay networks consisting of 
\textit{Edge Servers} (nodes) responsible for delivery of the contents from an original server to the end users~\cite{Clarck2006,Afergan2006}. Current commercial CDN's like \textit{Akamai} use path diversity based 
techniques like \textit{SureRoute} to ensure that the edge servers maintain reliable connections to the 
original server. Video server selection schemes are discussed in~\cite{Guo2003} to maximize path diversity 
in CDN's.

Moreover, references~\cite{Mao2005} and~\cite{Nguyen2003} study the
problem of rate allocation over multiple paths. Assuming each path
follows the leaky bucket model, reference~\cite{Mao2005} shows that
a water-filling scheme provides the minimum end-to-end delay. On the
other hand, reference~\cite{Nguyen2003} considers a scenario of
multiple senders and a single receiver, assuming all the senders
share the same source of data. The connection between each sender
and the receiver is assumed to follow the Gilbert-Elliot model. They
propose a receiver-driven protocol for packet partitioning and rate
allocation. The packet partitioning algorithm ensures no sender
sends the same packet, while the rate allocation algorithm minimizes
the probability of irrecoverable loss in the FEC
scheme~\cite{Nguyen2003}. They only address the rate allocation
problem for the case of two paths. A brute-force search algorithm is
proposed in~\cite{Nguyen2003} to solve the problem. Generalization
of this algorithm over multiple paths results in an exponential
complexity in terms of the number of paths. Moreover, it should be
noted that the scenario of~\cite{Nguyen2003} is equivalent, without
any loss of generality, to the case in which multiple independent
paths connect a pair of end-nodes as they assume the senders share
the same data.

\textit{Maximum Distance Separable} (MDS) codes have been shown to
be optimum in the sense that they achieve the maximum possible
minimum distance ($d_{min}$) among all the block codes of the same
size~\cite{Roth2006MDS}. Indeed, any $[N, K]$ MDS code (with block
length $N$ and $K$ information symbols) can be successfully
recovered from any subset of its entries of length $K$ or more. This
property makes MDS codes favorable FEC schemes over the erasure
channels like the Internet~\cite{Zakhor2001,Dairaine2005,Peng2005}.
However, the simple and practical encoding-decoding algorithms for
such codes have quadratic time complexity in terms of the code
size~\cite{Alon1995}. Theoretically, more efficient ($O\left(
N\log^2\left(N\right) \right)$) MDS codes can be constructed based
on evaluating and interpolating polynomials over specially chosen
finite fields using Discrete Fourier Transform~\cite{Justesen1976},
but these methods are not competitive in practice with the simpler
quadratic methods except for extremely large block sizes. Recently,
a family of almost-MDS codes with low encoding-decoding time
complexity (linear in term of the code length) is proposed and shown
to be practical over the erasure channels like the
Internet~\cite{Luby2001, Shokrollahi2006}. In these codes, any
subset of symbols of size $K(1+\epsilon)$ is sufficient to recover
the original $K$ symbols with high
probability~\cite{Shokrollahi2006}.

MDS codes also require alphabets of a large size. Indeed, all the
known MDS codes have alphabet sizes growing at least linearly with
the block length $N$. There is a conjecture stating that all the
$[N, K]$ MDS codes over the Galois field $\mathbb{F}_q$ with
$1<K<N-1$ have the property that $N \leq q+1$ with two exceptions
\cite{Roth2006MDS}. However, this is not an issue in the practical
networking applications since the alphabet size is $q=2^r$ where $r$
is the packet size, i.e. the block size is much smaller than the
alphabet size. Algebraic computation over Galois fields
($\mathbb{F}_q$) of such cardinalities is now practically possible
with the increasing processing power of electronic circuits. Note that network 
coding schemes, recently proposed and applied for content distribution over 
large networks, have a comparable computational 
complexity~\cite{Koetter2003, Chou2003, Gkantsidis2005}.

In this work, we utilize path diversity to improve the performance
of FEC between two end-nodes over a general packet switched network like the 
Internet. The details of path setup process is not discussed here. More precisely, it
is assumed that $L$ independent paths are set up by a smart overlay
network or any other means~\cite{Andersen2001,Chun2004,Han2005,Han2006,Han20062,Han2004,Guo2003,Akella2008,Akella2004,Andersen2003,Srinivasan2007,Cha2006,Nelakuditi2004}. Each path is modeled by a two-state continuous time Markov process called
Gilbert-Elliot channel~\cite{Bolot1999,Bolot1996,Nguyen2003,Nguyen2004,Leannec1999}. Probability of 
irrecoverable loss ($P_E$) is defined as the measure of FEC performance. It is known that MDS block codes
have the minimum probability of error over our \textit{End-to-End 
Channel} model, and over any other erasure channel with or without
memory~\cite{Fashandi2008isit,Fashandi20083}. Applying MDS
codes, our analysis shows an exponential decay of $P_E$ with respect
to $L$ for the asymptotic case where the number of paths is large.
Of course, in many practical cases, the number of \textit{disjoint}
or \textit{independent} paths between the end nodes is limitted.
However, in our asymptotic analysis, we have assumed that it is
possible to find $L$ independent paths between the end points even
when $L$ is large. Moreover, the optimal rate allocation problem is
solved in the asymptotic case. It is seen that in the asymptotically
optimal rate allocation, each path is assigned a positive rate
\textit{iff} its quality is above a certain threshold. Quality of a
path is defined as the percentage of the time it spends in the bad
state. Furthermore, using dynamic programming, a heuristic
suboptimal algorithm is proposed for rate allocation over a finite
number of paths (limitted $L$). Unlike the brute-force search, this algorithm has a
polynomial complexity, in terms of the number of paths. It is shown
that the result of this algorithm converges to the asymptotically
optimal solution for large number of paths. Finally, the proposed
algorithm is simulated and compared with the optimal rate allocation
found by exhaustive search for practical number of paths. Simulation
results verify the near-optimal performance of the proposed
suboptimal algorithm in practical scenarios.

The rest of this paper is organized as follows.
Section~\ref{section:SysModelProbState} describes the system model.
Probability distribution of the bad burst duration is discussed in
section~\ref{section:ProbDistBadBurst}. Performance of FEC in three
cases of a single path, multiple identical paths, and non-identical
paths are analyzed in section~\ref{section:MultiplePath}.
Section~\ref{section:SubOptimalRateAlloc} studies the rate
allocation problem, and proposes a suboptimal rate allocation
algorithm. Finally, section~\ref{section:Conclusion} concludes the
paper.

\begin{figure}
    \centering
    \includegraphics[scale=0.20]{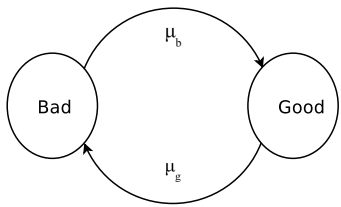}
    \caption{Continuous-time two-state Markov model of the end-to-end channel}
    \label{fig:GilbertCell}
\end{figure}

\section{System Modeling and Formulation}
\label{section:SysModelProbState}

\subsection{End-to-End Channel Model}
\label{subsection:InternetChModel}

From an end to end protocol's perspective, performance of the lower
layers in the protocol stack can be modeled as a random
\textit{channel} called the \textit{end-to-end channel}. Since each
packet usually includes an internal error detection coding (for
instance a Cyclic Redundancy Check), the end-to-end channel is satisfactorily modeled as
an erasure channel. Delay of the end-to-end channel is strongly
dependent on its packet loss pattern, and affects the QoS
considerably~\cite{Yajnik99, Rossi2003}.

In this work, the model assumed for the end-to-end channel is a
two-state Markov model called Gilbert-Elliot cell, depicted in
Fig.~\ref{fig:GilbertCell}. The channel spends an exponentially
distributed random amount of time with the mean $\frac{1}{\mu_{g}}$
in the \textit{Good} state. Then, it alternates to the \textit{Bad}
state and stays in that state for another random duration
exponentially distributed with the mean $\frac{1}{\mu_{b}}$. It is
assumed that the channel state does not change during the
transmission of a given packet~\cite{Nguyen2004,Kellerer2002,Henocq2000}. Hence, if a packet
is transmitted from the source at anytime during the good state, it
will be received correctly. Otherwise, if it is transmitted during
the bad state, it will eventually be lost before reaching the
destination. Therefore, the average probability of error is equal to
the steady state probability of being in the bad state, 
$\pi_{b}=\frac{\mu_{g}}{\mu_{g}+\mu_{b}}$. To have a reasonably low
probability of error, $\mu_{g}$ must be much smaller than $\mu_{b}$.
This model is widely used in the literature for theoretical analysis where delay is
not a significant factor~\cite{Bolot1999,Bolot1996,Nguyen2003,Nguyen2004,Leannec1999,Leannec19992,Kellerer2002,Henocq2000}. Despite its simplicity, this model satisfactorily captures the bursty error characteristic
of the end-to-end channel. More comprehensive models like the hidden
Markov model are introduced in~\cite{Rossi2003,Salamatian2001}. Although
analytically cumbersome, such models express the dependency of loss
and delay more accurately.

\begin{figure}
     \centering
     \subfigure[]{
           \label{fig:SSSD}
           \includegraphics[width=0.45\textwidth]{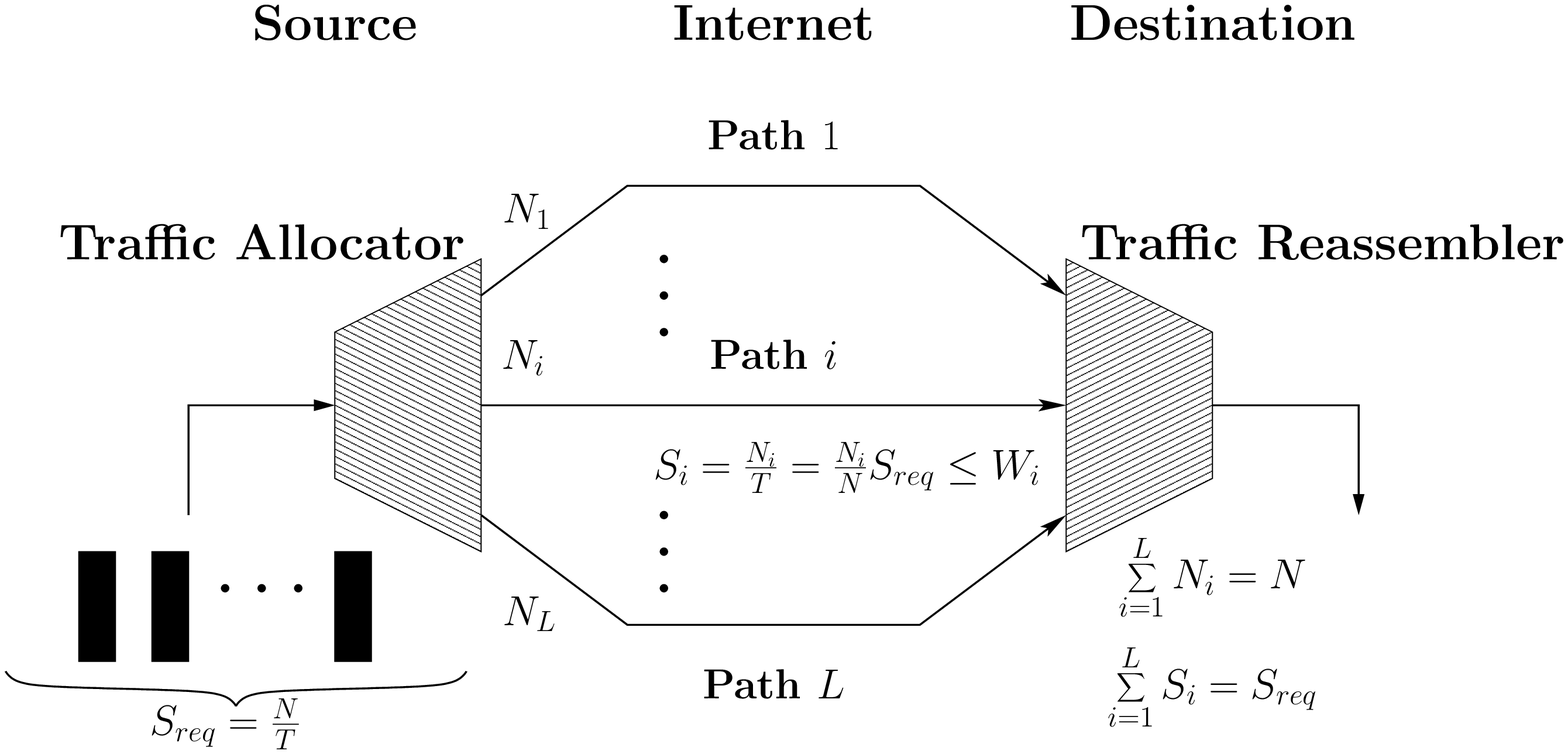}}
     \vspace{0.0in}
     \subfigure[]{
          \label{fig:SSSDsub}
          \includegraphics[width=0.45\textwidth]{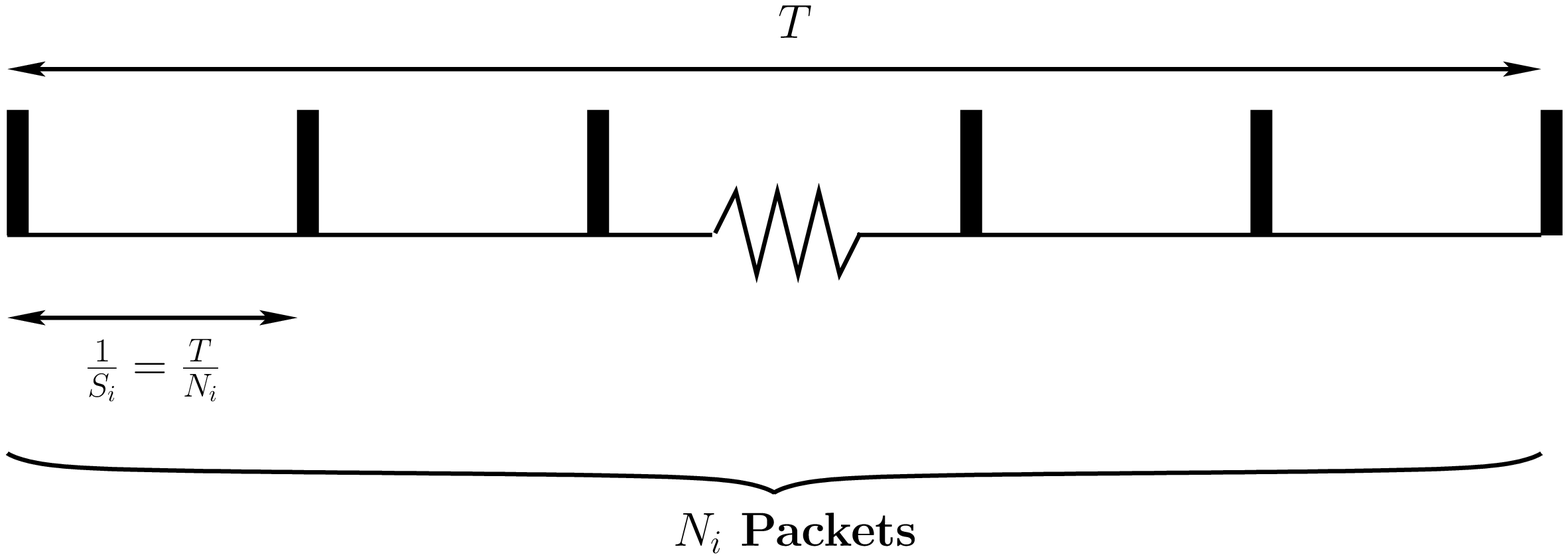}}
     \caption{Rate allocation problem: a block of $N$ packets is being sent
     from the source to the destination through $L$ independent paths over the
     network during the time interval $T$ with the required rate $S_{req}=\frac{N}{T}$.
     The block is distributed over the paths according to the vector $\mathbf{N}=(N_1,\dots,N_L)$
     which corresponds to the rate allocation vector $\mathbf{S}=(S_1,\dots,S_L)$ }
     \label{fig:SSSDmain}
\end{figure}

\subsection{Typical FEC Model}
\label{subsection:FECmodel} A concatenated coding is used for packet
transmission. The coding inside each packet can be a simple Cyclic Redundancy Check (CRC) 
which enables the receiver to detect an error inside each packet.
Then, the receiver can consider the end-to-end channel as an erasure
channel. Other than the coding inside each packet, a \textit{Forward
Error Correction} (FEC) scheme is applied between packets. Every $K$
packets are encoded to a \textit{Block} of $N$ packets where $N>K$
to create some redundancy. The $N$ packets of each block are
distributed across the $L$ available independent paths, and are
received at the destination with some loss (erasure). The ratio of
$\alpha=\frac{N-K}{N}$ defines the FEC overhead. A \textit{Maximum
Distance Separable} (MDS) $[N,K]$ code, such as the Reed-Solomon
code, can reconstruct the original $K$ data packets at the receiver
side if $K$ or more of the $N$ packets are received
correctly~\cite{Roth2006RSdec}. According to the following theorem, an MDS
code is the optimum block code we can design over any erasure
channel. Although FEC imposes some bandwidth overhead, it might be
the only option when feedback and retransmission are not feasible or
fast enough to provide the desirable QoS.

\textbf{Definition I.} An erasure channel is defined as the one which
maps every input symbol to either itself or to an erasure symbol
$\xi$. More accurately, an arbitrary channel (memoryless or with memory) with the input 
vector $\mathbf{x}\in\mathcal{X}^N$, $|\mathcal{X}|=q$ , the output 
vector $\mathbf{y}\in \left(\mathcal{X} \cup \{\xi\}\right)^N$, and the transition
probability $p\left(\mathbf{y}|\mathbf{x}\right)$ is defined to be erasure \textit{iff} it satisfies the 
following conditions:
\begin{enumerate}
\item $p\left( y_j \notin \left\{ x_j,\xi \right\} \right | x_j)=0,~\forall~j$.
\item Defining the erasure identifier vector $\mathbf e$ as
\begin{equation}
e_j = \left \{ \begin{array}{ll}
1 & y_j = \xi \\
0 & \mbox{otherwise} \end{array} \right. \nonumber
\end{equation}
$p(\mathbf{e} | \mathbf{x})$ is independent of $\mathbf x$.
\end{enumerate}

\textbf{Theorem I.}  A block code of size $[N,K]$ with equiprobable codewords over an arbitrary erasure 
channel (memoryless or with memory) has the minimum probability of error (assuming optimum, i.e.,
maximum likelihood decoding) among all block codes of the same size
\textit{if} that code is \textit{Maximum Distance Separable} (MDS).
The proof is given in~\cite{Fashandi2008isit,Fashandi20083}.

\subsection{Rate Allocation Problem}
\label{subsection:RateAllocProb} The network is modeled as follows.
$L$ independent paths, $1,2,\dots,L$, connect the source to the
destination, as indicated in Fig.~\ref{fig:SSSD}. Information bits
are transmitted as packets, each of a constant length $r$.
Furthermore, there is a constraint on the maximum rate for each path,
meaning that the $i$'th path can support a maximum rate of $W_i$
packets per second. This constraint can be considered as an
upperbound imposed by the physical characteristics of the path. As
an example,~\cite{Padhye2000} introduces the concept of the
\textit{maximum TCP-friendly bandwidth} for the maximum capacity of
an Internet path. $W_i$'s are assumed to be known at the transmitter
side. For a specific application and FEC scheme, we require a rate
of $S_{req}$ packets per second from the source to the destination.
Obviously, we should have $S_{req} \leq \sum_{i=1}^{L} W_{i}$ to
have a feasible solution. The information packets are assumed to be
coded in blocks of length $N$ packets. Hence, it takes
$T=\frac{N}{S_{req}}$ seconds to transmit a block of packets. In
practical scenarios with finite number of paths, the end-to-end
required rate ($S_{req}$) is given, and the values of $N$ and $T$
have to be chosen based on the feasible complexity of the MDS
decoder and the delay constraint of the application, respectively.

According to the FEC model, we can send $N_{i}$ packets through the
path $i$ as long as $\sum_{i=1}^{L} N_{i}=N$ and $\frac{N_i}{T}\leq
W_i$. The rate assigned to path $i$ can be expressed as
$S_{i}=\frac{N_i}{T}=\frac{N_i}{N}S_{req}$, since the transmission
instants of the $N_i$ packets are distributed evenly over the block
duration $T$ (see Fig.~\ref{fig:SSSDsub}). Obviously, we have
$\sum_{i=1}^{L} S_{i}=S_{req}$. The objective of rate allocation
problem is to find the optimal rate allocation vector or the vector
$\mathbf{N}=(N_1,\cdots,N_L)$ which minimizes the probability of
irrecoverable loss ($P_E$).

The above formulation of rate allocation problem is valid for any
finite number of paths and any chosen values of $N$ and $T$.
However, in section~\ref{section:MultiplePath} where the performance
of path diversity is studied for a large number of paths, and also
in Theorem III where the optimality of the proposed suboptimal
algorithm is proved for the asymptotic case, we assume that $N$
grows linearly in terms of the number of paths, i.e. $N=n_0 L$, for
a fixed $n_0$. The reason behind this assumption is that when $L$
grows asymptotically large, the number of paths eventually exceeds
the block length, if $N$ stays fixed. Thus, $L-N$ paths become
useless for the values of $N$ larger than $N$. At the same time, it
is assumed that the delay imposed by FEC, $T$, stays fixed with
respect to $L$. This model results in a linearly increasing rate as
the number of paths grows. We will later show that utilizing
multiple paths, it is possible to simultaneously achieve an
exponential decay in $P_E$ and a linear increase in rate, while the
delay stays constant.

In this work, an irrecoverable loss is defined as the event where
more than $N-K$ packets are lost in a block of $N$ packets. $P_E$
denotes the probability of this event. It should be noted that this
probability is different from the decoding error probability of a
maximum likelihood decoder performed on an MDS $[N,K]$ code, denoted
by $\mathbb{P}\{\mathcal{E}\}$.  Theoretically, an optimum maximum
likelihood decoder of an MDS code may still decode the original
codeword correctly with a positive, but very small probability, if
it receives less than $K$ symbols (packets). More precisely, such a
decoder is able to correctly decode an MDS code over $\mathbb{F}_q$
with the probability of $\frac{1}{q^i}$ after receiving $K-i$
correct symbols (see the proof of Theorem I in~\cite{Fashandi2008isit,Fashandi20083} for more details). Of
course, for Galois fields with a large cardinality, this probability
is usually negligible. The relationship between $P_E$ and
$\mathbb{P}\{\mathcal{E}\}$ can be summarized as follows:
\begin{eqnarray}
\mathbb{P}\{\mathcal{E}\} & = & P_E - \sum_{i=1}^{K} \dfrac{\mathbb{P\{\mbox{$K-i$ Packets received correctly}\}}}{q^i}    \nonumber \\
                          & \geq & P_E - \dfrac{1}{q} \sum_{i=1}^{K} \mathbb{P\{\mbox{$K-i$ Packets received correctly}\}} \nonumber \\
                          & = & P_E \left( 1-\dfrac{1}{q} \right).
\end{eqnarray}
Hence, $\mathbb{P}\{\mathcal{E}\}$ is bounded as
\begin{equation}
P_E \left( 1-\dfrac{1}{q} \right)  \leq  \mathbb{P}\{\mathcal{E}\}  \leq  P_E.
\label{equation:PEPEboundedML}
\end{equation}

The reason $P_E$ is used as the measure of system performance is 
that while many practical low-complexity decoders for MDS codes
work perfectly if the number of correctly received symbols is at
least $K$, their probability of correct decoding is much less than
that of maximum likelihood decoders when the number of correctly
received symbols is less than $K$~\cite{Roth2006RSdec}. Thus, in the rest of this paper,
$P_E$ is used as a close approximation of decoding error.


\section{Probability Distribution of Bad Bursts}
\label{section:ProbDistBadBurst}

The continuous random variable $B_i$ is defined as the duration of
time that the path $i$ spends in the bad state in a block duration,
$T$. We denote the values of $B_{i}$ with parameter $t$ to emphasize
that they are expressed in the unit of time. In this section, we
focus on one path, for example path 1. Therefore, the index $i$ can
be temporarily dropped in analyzing the probability distribution
function (pdf) of $B_{i}$.

We define the events $g$ and $b$, respectively, as the channel being
in the good or bad states at the start of a block. Then, the
distribution of $B$ can be written as
\begin{equation}
f_{B}(t)=f_{B|b}(t)\pi_{b}+f_{B|g}\pi_{g}.
\label{equation:f_Bt}
\end{equation}
To proceed further, two assumptions are made. First, it is assumed
that $\pi_{g}\gg \pi_{b}$ or equivalently
$\frac{1}{\mu_{g}}\gg\frac{1}{\mu_{b}}$. This condition is valid for
a channel with a reasonable quality. Besides, the block time $T$ is
assumed to be much shorter than the average good state duration
$\frac{1}{\mu_{g}}$, i.e. $1\gg\mu_{g}T$, such that $T$ can contain
either none or a single interval of bad
burst~(see~\cite{Nguyen2003,Nguyen2004,Bolot1999} for
justification). More precisely, the probability of having at least
two bad bursts is negligible compared to the probability of having
exactly one bad burst. However, it should be noted that all the
results of this paper except subsection \ref{subsection:SinglePath}
remain valid regardless of these two assumptions. Of course, in that
case, the exact probability distribution function of $B_i$ should be
used instead of the approximation used here (refer to Remark I in
subsection \ref{subsection:IdenticalPaths}).

Hence, the pdf of $B$ conditioned on the event $b$ can be approximated as
\begin{equation}
f_{B|b}(t)=\mu_{b}e^{-\mu_{b}t}+\delta(t-T)e^{-\mu_{b}T}
\label{equation:f_Bbt}
\end{equation}
where $\delta(u)$ is the Dirac delta
function.~\eqref{equation:f_Bbt} follows from the memoryless nature
of the exponential distribution, the assumption that $T$ contains at
most one bad burst, and the fact that any bad burst longer than $T$
has to be truncated at $B=T$.

To compute $f_{B|g}(t)$, we have
\begin{equation}
f_{B|g}(t)=\mathbb{P}\{B=0|g\}\delta(t)-\frac{\partial}{\partial t}\mathbb{P}\{B>t|g\}
\label{equation:f_Bgt}
\end{equation}
where
\begin{equation}
\mathbb{P}\{B=0|g\}=e^{-\mu_{g}T}\approx 1-\mu_{g}T
\label{equation:PB0g}
\end{equation}
and
\begin{equation}
\mathbb{P}\{B>t|g\} \stackrel{(a)}{=} (1-e^{-\mu_{g}(T-t)})e^{-\mu_{b}t} \approx \mu_{g}(T-t)e^{-\mu_{b}t}
\label{equation:PBtg}
\end{equation}
where $(a)$ results from the fact that $\{B>t|g\}$ is equivalent to
the initial good burst being shorter than $T-t$, and the following
bad burst larger than $t$, and the duration $T$ containing at most
one bad burst. Now,
combining~\eqref{equation:f_Bbt},~\eqref{equation:f_Bgt},~\eqref{equation:PB0g},
and~\eqref{equation:PBtg}, $f_B(t)$ can be computed.

\begin{figure}
    \centering
    \includegraphics[scale=0.14]{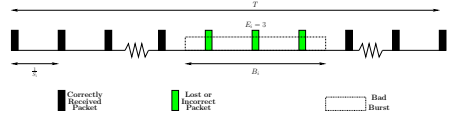}
    \caption{A bad burst of duration $B_i$ happens in a block of length $T$. $E_i=3$ packets are
corrupted or lost during the interval $B_i$. Packets are transmitted every $\frac{1}{S_i}$ seconds, where
$S_i$ is the rate of path $i$ in $pkt/sec$.}
    \label{fig:Disc2Cont}
\end{figure}

%

\subsection{Discrete to Continuous Approximation}
\label{subsection:Disc2ContConvert} To compute the probability of
irrecoverable loss ($P_E$), we have to find the probability of $k_i$
packets being lost out of the $N_i$ packets transmitted through the
path $i$, for $i$ from $1$ to $L$ and $k_i$ from $0$ to $N_i$. Let
us denote the number of erroneous or lost packets over the path $i$
with the random variable $E_i$. Any two subsequent packets
transmitted over the path $i$ are $\frac{1}{S_i}$ seconds apart in
time, where $S_i$ is the transmission rate over the $i$'th path. We
observe that the probability $\mathbb{P}\{E_i\geq k_i\}$ can be
approximated with the continuous counterpart
$\mathbb{P}\{B_i\geq\frac{k_i}{S_i}\}$ when the inter-packet
interval is much shorter than the typical bad burst ($\frac{1}{S_i}
\ll \frac{1}{\mu_b}$, or equivalently $\mu_b \ll S_i$). The
necessity of this condition can be intuitively justified as follows.
In case this condition does not hold, any two consecutive packets
have to be transmitted on two independent states of the channel.
Thus, no gain would be achieved by applying diversity over multiple
independent paths. Figure~\ref{fig:Disc2Cont} shows an example of
this approximation in detail. The continuous approximation
simplifies the mathematical analysis as discussed in
section~\ref{section:MultiplePath}.

%
\begin{figure}[t]
    \centering
    \includegraphics[scale=0.50]{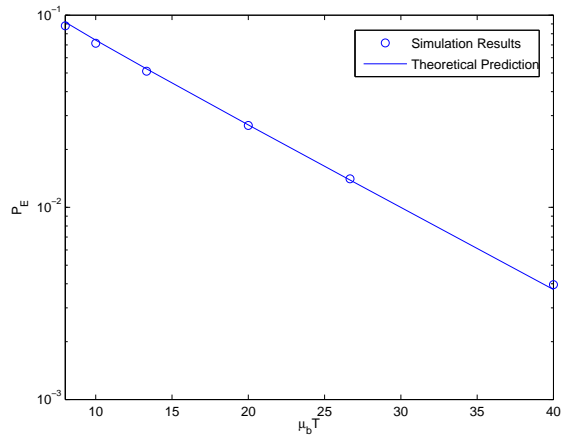}
    \caption{Probability of irrecoverable loss versus $\mu_{b}T$ for one path with fixed $\mu_{g}$, $T$ and $\alpha$.}
    \label{fig:PE1Path}
\end{figure}

\section{Performance Analysis of FEC on Multiple Paths}
\label{section:MultiplePath} Assume that a rate allocation algorithm
assigns $N_i$ packets to the path $i$. According to the discrete to continuous approximation in 
subsection~\ref{subsection:Disc2ContConvert}, when the $N_{i}$ packets of
the FEC block are sent over path $i$, the loss count can be
written as $\frac{B_{i}}{T}N_{i}$. Hence, the total ratio of
lost packets is equal to
$$
\sum_{i=1}^{L}\frac{B_{i}N_{i}}{TN}=\sum_{i=1}^{L}\frac{B_{i}\rho_{i}}{T}
$$
where $\rho_{i}=\frac{S_{i}}{S_{req}},~0 \leq \rho_i \leq 1$,
denotes the portion of the  bandwidth assigned to path $i$.
$x_{i}=\frac{B_{i}}{T}$ is defined as the portion of time that path
$i$ has been in the bad state ($0\leq x_{i}\leq 1$). Hence, the
probability of irrecoverable loss for an MDS code is equal to
\begin{equation}
P_{E}=\mathbb{P}\left\{\sum_{i=1}^{L}\rho_{i}x_{i}>\alpha\right\}
\label{equation:PErhoixi}
\end{equation}
where $\alpha=\frac{N-K}{N}$. In order to find the optimum rate
allocation, $P_E$ has to be minimized with respect to the allocation
vector ($\rho_i$'s), subject to the following constraints:
\begin{eqnarray}
0\leq\rho_{i}\leq \min\left\{1, \dfrac{W_i}{S_{req}}\right\},&\sum_{i=1}^{L}\rho_{i}=1
\label{equation:optimization}
\end{eqnarray}
where $W_i$ is the bandwidth constraint on path $i$ defined in
subsection~\ref{subsection:RateAllocProb}. Note that since $x_{i}$'s
are proportional to $B_i$'s, their pdf can be easily computed based
on the pdf of $B_i$'s.

\subsection{Performance of FEC on a Single Path}
\label{subsection:SinglePath}

Probability of irrecoverable loss for one path is equal to
\begin{equation}
P_{E}=\mathbb{P}\{B>\alpha T\}=\mathbb{P}\{B>\alpha T|b\}\pi_{b}+\mathbb{P}\{B>\alpha T|g\}\pi_{g} \nonumber
\end{equation}
where $\mathbb{P}\{B>\alpha T|b\}$ and $\mathbb{P}\{B>\alpha T|g\}$ can be computed as
\begin{eqnarray}
\mathbb{P}\{B>\alpha T|b\} & = \int_{\alpha T}^{T}f_{B|b}(t)dt & =e^{-\mu_{b}\alpha T}, \nonumber \\
\mathbb{P}\{B>\alpha T|g\} & = \int_{\alpha T}^{T}f_{B|g}(t)dt & =\mu_{g}(1-\alpha)Te^{-\mu_{b}\alpha T} \nonumber
\end{eqnarray}
when the assumptions in section~\ref{section:ProbDistBadBurst} and
equations~\eqref{equation:f_Bbt} and~\eqref{equation:PBtg} are used.
Thus, we have
\begin{eqnarray}
P_{E} & = & \pi_{b}e^{-\mu_{b}\alpha T}(1+\mu_{b}(1-\alpha)T) \nonumber \\
 & \stackrel{(a)}{\approx} & \left[ \frac{1}{\mu_b} + \left( 1 - \alpha \right) T \right]\mu_g e^{-\mu_{b}\alpha T}
\label{equation:Perr1Path}
\end{eqnarray}
where $(a)$ follows from the assumption that the end-to-end channel
has a low probability of error ($\frac{1}{\mu_g} \gg
\frac{1}{\mu_b}$).

As we observe, for large values of $\mu_b T$, $P_E$ decays
exponentially with $\mu_b T$. Figure~\ref{fig:PE1Path} shows the
results of simulating a typical scenario of streaming data between
two end-points with the rate $S_{req}=1000\frac{pkt}{sec}$, the
block length $N=200$, and the number of information packets $K=180$.
These values result in a block transmission time of $T=200ms$. The
average good burst of the end-to-end channel, $\mu_g$, is selected
such that $\mu_g T=\frac{1}{5}$. However, the average bad burst,
$\mu_b$, varies such that $\mu_b T$ varies from $8$ to $40$, in
accordance with the values in~\cite{Nguyen2003, Nguyen2004}. The
slope of the best linear fit (in semilog scale) to the simulation
points is $0.097$ which is in accordance with the value of $0.100$,
resulted from the theoretical
approximation in~\eqref{equation:Perr1Path}.


\subsection{Identical Paths}
\label{subsection:IdenticalPaths} 
When the paths are identical and have equal bandwidth 
constraints\footnote{The case where $W_i$'s are different is discussed in Remark V of subsection~\ref{subsection:NonIdenticalPaths}} ($W_i=W$ for $\forall~1\leq i \leq L$), due
to the symmetry of the problem, the uniform rate allocation
($\rho_{i}=\frac{1}{L}$) is obviously the optimum solution. Of
course, the solution is feasible only when we have $\frac{1}{L}\leq \frac{W}{S_{req}}$. Then, the probability of
irrecoverable loss can be simplified as
\begin{equation}
P_{E}=\mathbb{P}\left \{\frac{1}{L}\sum_{i=1}^{L}x_{i}>\alpha \right \}.
\label{equation:PEinequality}
\end{equation}
Let us define $Q(x)$ as the probability distribution function of
$x$. Since $x$ is defined as $x=\frac{B}{T}$, clearly we have
$Q(x)=Tf_B(xT)$. Defining $\mathbb{E}\{\}$ as the expected value
operator throughout this paper, $\mathbb{E}\{x\}$ can be computed
based on $Q(x)$. We observe that in (\ref{equation:PEinequality}),
the random variable $x_i$'s are bounded and independent. Hence, the
following well-known upperbound in large deviation
theory~\cite{Dembo1998} can be applied
\begin{eqnarray}
& P_{E}\leq e^{-u(\alpha)L}  & \nonumber \\
& u(\alpha)=\left\{ \begin{array}{ll}
           0 & \mbox{for $\alpha \leq \mathbb{E}\{x\}$}\\
           \lambda\alpha-\log(\mathbb{E}\{e^{\lambda x}\}) & \mbox{otherwise}\end{array} \right. &
\label{equation:ChernoffBound}
\end{eqnarray}
where the $\log$ function is computed in Neperian base, and
$\lambda$ is the solution of the following non-linear equation,
which is shown to be unique by Lemma I.
\begin{equation}
\alpha=\frac  {\mathbb{E}\{xe^{\lambda x}\}} {\mathbb{E}\{e^{\lambda x}\}}.
\label{equation:AlphaLambda}
\end{equation}
Since $\lambda$ is unique, we can define $l(\alpha)=\lambda$. Even
though being an upperbound,
inequality~\eqref{equation:ChernoffBound} is exponentially tight for
large values of $L$~\cite{Dembo1998}. More precisely
\begin{equation}
P_{E}\doteq e^{-u(\alpha)L}
\label{equation:dotEQmain}
\end{equation}
where the notation $\doteq$ means
$\displaystyle\lim_{L->\infty}-\frac{\log P_{E}}{L}=u(\alpha)$. Now,
we state two useful lemmas whose proofs can be found in the
appendices~\ref{section:ProofLemmaI} and~\ref{section:ProofLemmaII}.

\textbf{Lemma I.} $u(\alpha)$ and $l(\alpha)$ have the following properties:
\begin{enumerate}
\item $\frac{\partial}{\partial \alpha}l(\alpha)>0$
\item $l \left( \alpha=0 \right)=-\infty$
\item $l \left( \alpha=\mathbb{E}\{x\} \right)=0$
\item $l \left( \alpha=1 \right) =+\infty$
\item $\frac{\partial}{\partial \alpha}u(\alpha)=l(\alpha)>0$ for $\alpha>\mathbb{E}\{x\}$
\end{enumerate}

\textbf{Lemma II.} Defining $y=\frac {1} {L} \sum_{i=1}^{L}{x_i}$,
where $x_i$'s are i.i.d. random variables as already defined, the
probability density function of $y$ satisfies $f_y(\alpha) \doteq
e^{-u(\alpha) L}$, for all $\alpha > \mathbb{E}\{x\}$.

\begin{figure}
     \centering
     \subfigure[]{
           \label{fig:PEmultiPath}
           \includegraphics[width=0.45\textwidth]{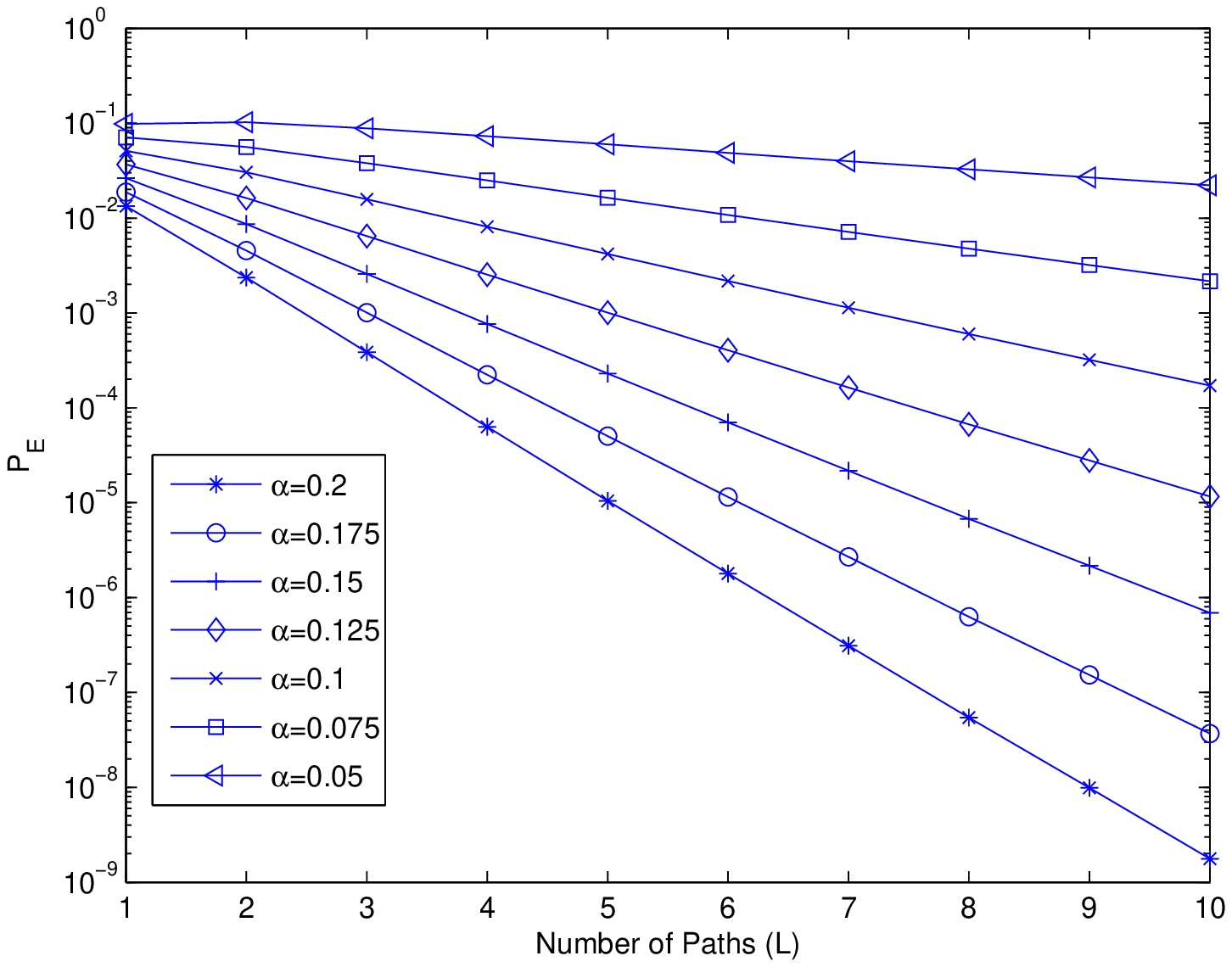}}
     \vspace{0.0in}
     \subfigure[]{
          \label{fig:ExpMultiPath}
          \includegraphics[width=0.45\textwidth]{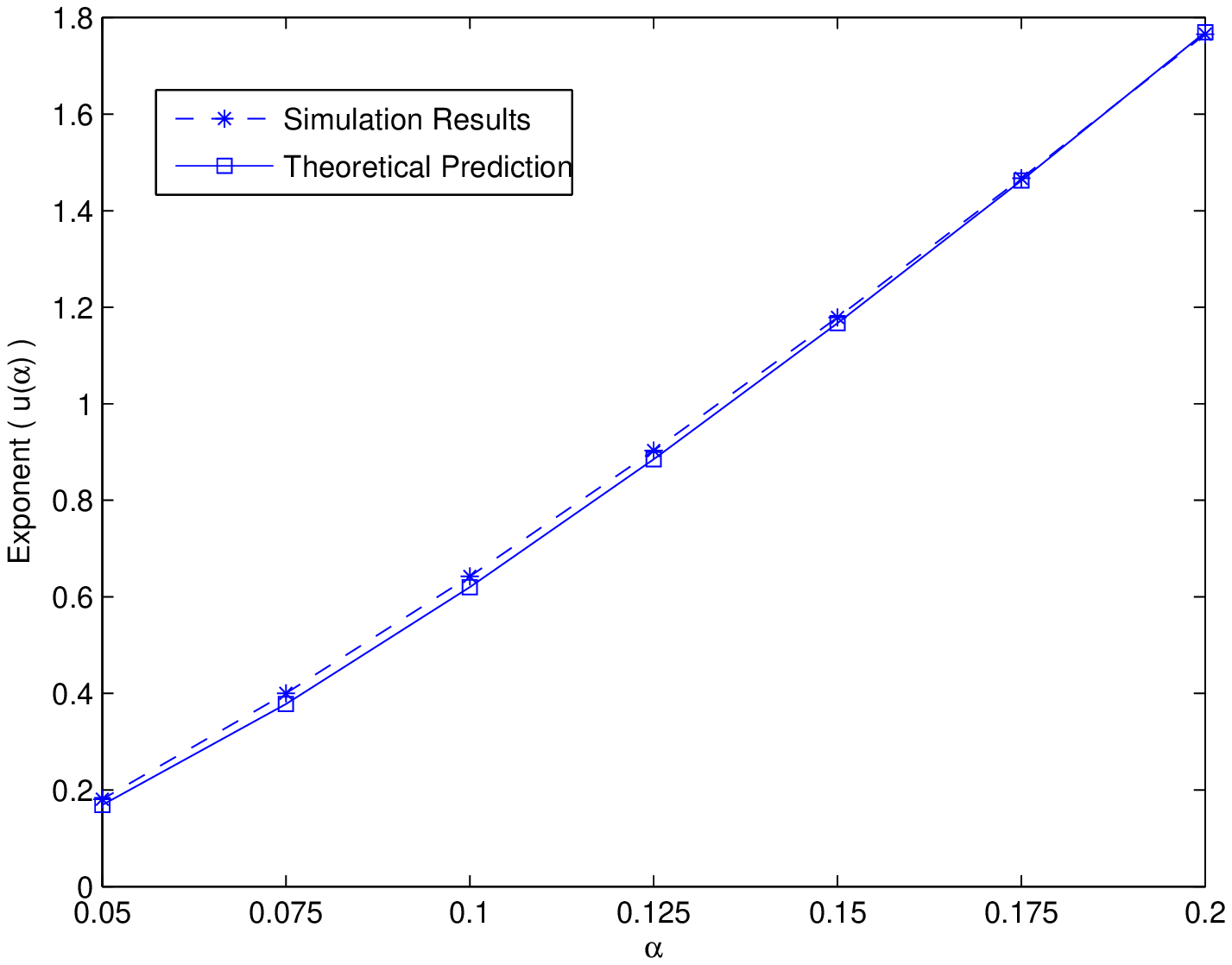}}
     \caption{(a) $P_E$ vs. $L$ for different values of $\alpha$. (b) The exponent (slope) of plot (a) for different values of $\alpha$: experimental versus theoretical values.}
     \label{fig:TwoPlots}
\end{figure}

Figure~\ref{fig:TwoPlots} compares the theoretical and simulation
results. We assume the block transmission time is $T=200ms$. The
block length is proportional to the number of paths as $N=20L$. The
average good burst of the end-to-end channel, $\mu_g$, is selected
such that $\mu_g T=\frac{1}{5}$. The end-to-end channel has the error
probability of $\pi_b=0.015$. Coding overhead is changed from
$\alpha=0.05$ to $\alpha=0.2$. The probability of irrecoverable loss
is plotted versus the number of paths, $L$, in semilogarithmic scale
in Fig.~\ref{fig:PEmultiPath} for different values of $\alpha$. We
observe that as $L$ increases, $\log P_E$ decays linearly which is
expected noting equation~\eqref{equation:ChernoffBound}. Also,
Fig.~\ref{fig:ExpMultiPath} compares the slope of each plot in
Fig.~\ref{fig:PEmultiPath} with $u(\alpha)$.
Figure~\ref{fig:TwoPlots} shows a good agreement between the theory
and the simulation results, and also verifies the fact that the
stronger the FEC code is (larger $\alpha$), the higher is the gain
we achieve through path diversity (larger exponent).


\textbf{Remark I.} Equation~\eqref{equation:dotEQmain} is a direct
result of the discrete to continuous approximation in
subsection~\ref{subsection:Disc2ContConvert}. Therefore, it remains
valid even if the other approximations in
section~\ref{section:ProbDistBadBurst} do not hold. For example, if
the block time contains more than one bad burst,
equations~\eqref{equation:f_Bbt} and~\eqref{equation:PBtg} are no
longer valid. However, equation~\eqref{equation:dotEQmain} is still
valid as long as the discrete to continuous approximation is used.
Of course, in this case, the exact distributions of $B$ and $x$
should be used to compute $u(\alpha)$ and $\lambda$ instead of their
simplified versions.

\textbf{Remark II.} A special case is when the block code uses all
the bandwidth of the paths. In this case, we have $N=LWT$, where $W$
is the maximum bandwidth of each path, and $T$ is the block
duration. Assuming $\alpha > \mathbb{E}\{x\}$ is a constant
independent of $L$, we observe that the information packet rate is
equal to $\frac{K}{T}=\left( 1 - \alpha \right)WL$, and the error
probability is $P_E \doteq e^{-u \left( \alpha \right) L}$. This
shows using MDS codes over multiple independent paths provides an
exponential decay in the irrecoverable loss probability and a
linearly growing end-to-end rate in terms of the number of paths,
simultaneously.


\subsection{Non-Identical Paths}
\label{subsection:NonIdenticalPaths}

Now, let us assume there are $J$ types of paths between the source
and the destination, consisting of $L_j$ identical paths of type $j$
($\sum_{j=1}^{J}{L_j}=L$). Without loss of generality, we assume
that the paths are ordered according to their associated type, i.e.
the paths from $1+\sum_{k=1}^{j-1}{L_k}$ to $\sum_{k=1}^{j}{L_k}$
are of type $j$. We denote $\gamma_j=\frac {L_j} {L}$. According to
the i.i.d. assumption, it is obvious that $\rho_i$ has to be the
same for all paths of the same type. $\eta_j$ and $y_j$ are defined
as
\begin{eqnarray}
\eta_j & =  & \sum_{\sum_{k=1}^{j-1}{L_k}<i\leq\sum_{k=1}^{j}{L_k}}{\rho_i}\nonumber \\
y_j    & =  & \frac {\eta_j} {L\gamma_j} \sum_{\sum_{k=1}^{j-1}{L_k}<i\leq\sum_{k=1}^{j}{L_k}}{x_i} .
\end{eqnarray}
Following Lemma II, we observe that $f_{y_j}(\beta_j) \doteq
e^{-\gamma_j u_j(\frac {\beta_j} {\eta_j}) L}$. We define the sets
$\mathcal{S}_I$, $\mathcal{S}_O$ and $\mathcal{S}_T$ as

\begin{eqnarray}
\mathcal{S}_I&=&\left\{ \left(\beta_1,\beta_2,\cdots,\beta_J\right)  | 0 \leq \beta_j\leq 1,\mbox{ } \sum_{j=1}^{J}{\beta_j}>\alpha \right\}\nonumber\\
\mathcal{S}_O&=&\left\{ \left(\beta_1,\beta_2,\cdots,\beta_J\right)  | 0 \leq \beta_j\leq 1,\mbox{ } \sum_{j=1}^{J}{\beta_j}=\alpha \right\} \nonumber \\
\mathcal{S}_T&=&\left\{ \left(\beta_1,\beta_2,\cdots,\beta_J\right)  | \eta_j \mathbb{E} \left\{ x_j \right\} \leq \beta_j  , \sum_{j=1}^{J}{\beta_j}=\alpha \right\} \nonumber
\end{eqnarray}
respectively. Hence, $P_E$ can be written as
\begin{eqnarray}
P_E & = & \mathbb{P} \left\{ \displaystyle\sum_{j=1}^{J} y_j > \alpha \right\} \nonumber \\
    & = & \int_{\mathcal{S}_I}{\displaystyle\prod_{j=1}^{J}{f_{y_j}(\beta_j)d \beta_j} } \nonumber \\
    & \doteq & \int_{\mathcal{S}_I}{e^{-L\displaystyle\sum_{j=1}^{J}{\gamma_j u_j(\frac {\beta_j} {\eta_j})}}d \beta_j } \nonumber \\
    & \stackrel{(a)}{\doteq} & e^{-L \displaystyle\min_{\boldsymbol{\beta} \in \mathcal{S}_I \cup \mathcal{S}_O}{ \displaystyle\sum_{j=1}^{J}{\gamma_j u_j\left(\frac {\beta_j} {\eta_j}\right)} }} \nonumber \\
    & \stackrel{(b)}{\doteq} & e^{-L \displaystyle\min_{\boldsymbol{\beta} \in \mathcal{S}_O}{ \displaystyle\sum_{j=1}^{J}{\gamma_j u_j\left(\frac {\beta_j} {\eta_j}\right)} }} \nonumber \\
    & \stackrel{(c)}{\doteq} & e^{-L \displaystyle\min_{\boldsymbol{\beta} \in \mathcal{S}_T}{ \displaystyle\sum_{j=1}^{J}{\gamma_j u_j\left(\frac {\beta_j} {\eta_j}\right)} }}  \nonumber \\
    & \stackrel{(d)}{\doteq} & e^{-L \displaystyle\sum_{j=1}^{J}{\gamma_j u_j\left(\frac {\beta_j^{\star}} {\eta_j}\right)}  } \label{equation:betastar}
\end{eqnarray}
where $(a)$ follows from Lemma III, $(b)$ follows from the fact that
$u_j(\alpha)$ is a strictly increasing function of $\alpha$, for
$\alpha > \mathbb{E}\{x_j\}$, and $(c)$ can be proved as follows.
Let us denote the vector which minimizes the exponent over the set
$\mathcal{S}_O$ as $\hat{\boldsymbol\beta}^\star$. Since
$\mathcal{S}_T$ is a subset of $\mathcal{S}_O$,
$\hat{\boldsymbol\beta}^\star$ is either in $\mathcal{S}_T$ or in
$\mathcal{S}_O-\mathcal{S}_T$. In the former case, $(c)$ is
obviously valid. When $\hat{\boldsymbol\beta}^\star \in
\mathcal{S}_O-\mathcal{S}_T$, we can prove that $0 \leq
\hat{\beta}_j^\star \leq \eta_j \mathbb{E}\{x_j\}$, for all $1 \leq
j \leq J$, by contradiction. Let us assume the opposite is true,
i.e., there is at least one index $1 \leq j \leq J$ such that $0
\leq \hat{\beta}_j^\star \leq \eta_j \mathbb{E}\{x_j\}$, and at
least one other index $1 \leq k \leq J$ such that $\eta_k
\mathbb{E}\{x_k\} < \hat{\beta}_k^\star$. Then, knowing that the
derivative of of $u_j(\alpha)$ is zero for
$\alpha=\mathbb{E}\{x_j\}$ and strictly positive for
$\alpha>\mathbb{E}\{x_j\}$, a small increase in
$\hat{\beta}_j^\star$ and an equal decrease in $\hat{\beta}_k^\star$
reduces the objective function, $\sum_{j=1}^{J}{\gamma_j
u_j\left(\frac {\beta_j} {\eta_j}\right)}$, which contradicts the
assumption that $\hat{\boldsymbol\beta}^\star$ is a minimum point.
Knowing that $0 \leq \hat{\beta}_j^\star < \eta_j
\mathbb{E}\{x_j\}$, for all $1 \leq j \leq J$, it is easy to show
that the minimum value of the objective function is zero over
$\mathcal{S}_O$, and $\mathcal{S}_T$ has to be an empty set.
Defining the minimum value of the positive objective function as
zero over an empty set ($\mathcal{S}_T$) makes $(c)$ valid for the
latter case where $\hat{\boldsymbol\beta}^\star \in
\mathcal{S}_O-\mathcal{S}_T$. Finally, applying Lemma IV results in
$(d)$ where $\boldsymbol \beta ^ \star$ is defined in the Lemma.

\textbf{Lemma III.} For any continuous positive function
$h(\mathbf{x})$ over a convex set $\mathcal{S}$, and defining $H(L)$
as
\begin{equation}
H(L)=\int_{\mathcal{S}}{e^{-h(\mathbf{x})L}d\mathbf{x}} \nonumber
\end{equation}
we have
\begin{equation}
\lim_{L\to\infty}-\frac{\log(H(L))}{L}=\inf_{\mathcal{S}}{h(\mathbf{x})}=\min_{cl(\mathcal{S})}{h(\mathbf{x})} \nonumber
\end{equation}
where $cl(\mathcal{S})$ denotes the closure of $\mathcal{S}$ (refer
to~\cite{Kelley1975} for the definition of the closure operator).
Proof of Lemma III can be found in
appendix~\ref{section:ProofLemmaIII}.

\textbf{Lemma IV.} There exists a unique vector
$\boldsymbol{\beta}^{\star}$ with the elements
$\beta_j^{\star}=\eta_j l_j^{-1} \left( \frac{\nu \eta_j}{\gamma_j}
\right)$ which minimizes the convex function
$\sum_{j=1}^{J}{\gamma_j u_j(\frac {\beta_j} {\eta_j})}$ over the
convex set $\mathcal{S}_T$, where $\nu$ satisfies the following
condition
\begin{equation}
\sum_{j=1}^{J}{\eta_j l_j^{-1}  \left( \dfrac{\nu \eta_j}{\gamma_j}  \right)  } = \alpha .
\label{equation:NuDef}
\end{equation}
$l^{-1}()$ denotes the inverse of the function $l()$ defined in
subsection~\ref{subsection:IdenticalPaths}. Proof of Lemma IV can be
found in appendix~\ref{section:ProofLemmaIV}.

Equation \eqref{equation:betastar} is valid for any fixed value of
$\boldsymbol{\eta}$. To achieve the most rapid decay of $P_E$, the
exponent must be maximized over $\boldsymbol{\eta}$.
\begin{equation}
\lim_{L\to\infty}-\frac{\log P_{E}}{L}=\max_{0\leq\eta_j\leq 1}{\sum_{j=1}^{J}{\gamma_j u_j\left(\frac {\beta_j^{\star}} {\eta_j}\right)}}
\label{equation:MaxEttaMinBeta}
\end{equation}
where $\boldsymbol{\beta}^{\star}$ is defined for any value of the
vector $\boldsymbol{\eta}$ in Lemma IV. Theorem II solves the
maximization problem in~\eqref{equation:MaxEttaMinBeta} and
identifies the asymptotically optimum rate allocation (for large
number of paths).

\textbf{Theorem II.} Consider a point-to-point connection over the
network with $L$ independent paths from the source to the
destination, each modeled as a Gilbert-Elliot cell, with a large
enough bandwidth constraint\footnote{By the term `large enough', we
mean the bandwidth constraint on a path of type $j$, $W_j$,
satisfies the condition $\frac{\eta_j n_0 }{T \gamma_j} \leq W_j$.
The reason is that $\eta_j$ must satisfy both conditions of $0 \leq
\eta_j \leq 1$ and $\frac{N_j}{TL_j}=\frac{\eta_j n_0 L}{T \gamma_j
L} \leq W_j$, simultaneously. When $W_j$ is large enough such that
$\frac{\eta_j n_0 }{T \gamma_j} \leq W_j$, the latter condition is
automatically satisfied, and the optimization problem can be
solved.}. The paths are from $J$ different types, $L_j$ paths from
the type $j$. Assume a block FEC of size $[N,K]$ is sent during a
time interval $T$. Let $N_j$ denote the number of packets in a block
of size $N$ assigned to the paths of type $j$, such that
$\sum_{j=1}^{J}N_j=N$. The rate allocation vector $\boldsymbol \eta$
is defined as $\eta_j=\frac{N_j}{N}$. For fixed values of
$\gamma_j=\frac{L_j}{L},~n_0=\frac{N}{L},~k_0=\frac{K}{L},~T$ and
asymptotically large number of paths $L$, the optimum rate
allocation vector $\boldsymbol{\eta}^{\star}$ can be found by solving the
following optimization problem:
\begin{eqnarray}
\max_{\boldsymbol{\eta}} g(\boldsymbol{\eta}) , \nonumber \\
\mbox{s.t. }\sum_{j=1}^{J}{\eta_j} = 1,~0 \leq \eta_j \leq 1 \nonumber
\end{eqnarray}
where $g(\boldsymbol{\eta})= \sum_{j=1}^{J} \gamma_j u_j\left(\frac
{\beta_j^{\star}} {\eta_j}\right)$, and $\boldsymbol{\beta}^{\star}$
is an implicit function of $\boldsymbol{\eta}$ defined in Lemma IV.
The functions $u_j()$ and $l_j()$ are defined in
subsections~\ref{subsection:IdenticalPaths}
and~\ref{subsection:NonIdenticalPaths}. Solving the above
optimization problem gives the unique solution
$\boldsymbol{\eta}^{\star}$ as
\begin{equation}
\eta_j^{\star}= \left\{ \begin{array}{ll}
           0 & \mbox{if $\alpha \leq \mathbb{E}\{x_j\}$}\\
             & \\
           \dfrac{ \gamma_j l_j(\alpha) } { \displaystyle\sum_{i=1,~\alpha > \mathbb{E}\{x_i\}}^{J} \gamma_i l_i(\alpha) } & \mbox{otherwise}\end{array} \right.
\label{equation:etaStar}
\end{equation}
if there is at least one $1 \leq j \leq J$ for which $\alpha >
\mathbb{E}\{x_j\}$. Otherwise, when $\alpha \leq \mathbb{E}\{x_j\}$
for all $1 \leq j \leq J$, the maximum value is zero for any
arbitrary rate allocation vector, $\boldsymbol{\eta}$. In any case,
the maximum value of the objective function is
$g(\boldsymbol{\eta}^{\star})=\sum_{j=1}^{J}\gamma_j u_j(\alpha)$
which is indeed the exponent of $P_E$ versus $L$. The proof of the
theorem can be found in appendix \ref{section:ProofTheoremII}.

\begin{figure}
     \centering
     \subfigure[]{
           \label{fig:TwoTypes}
           \includegraphics[width=0.45\textwidth]{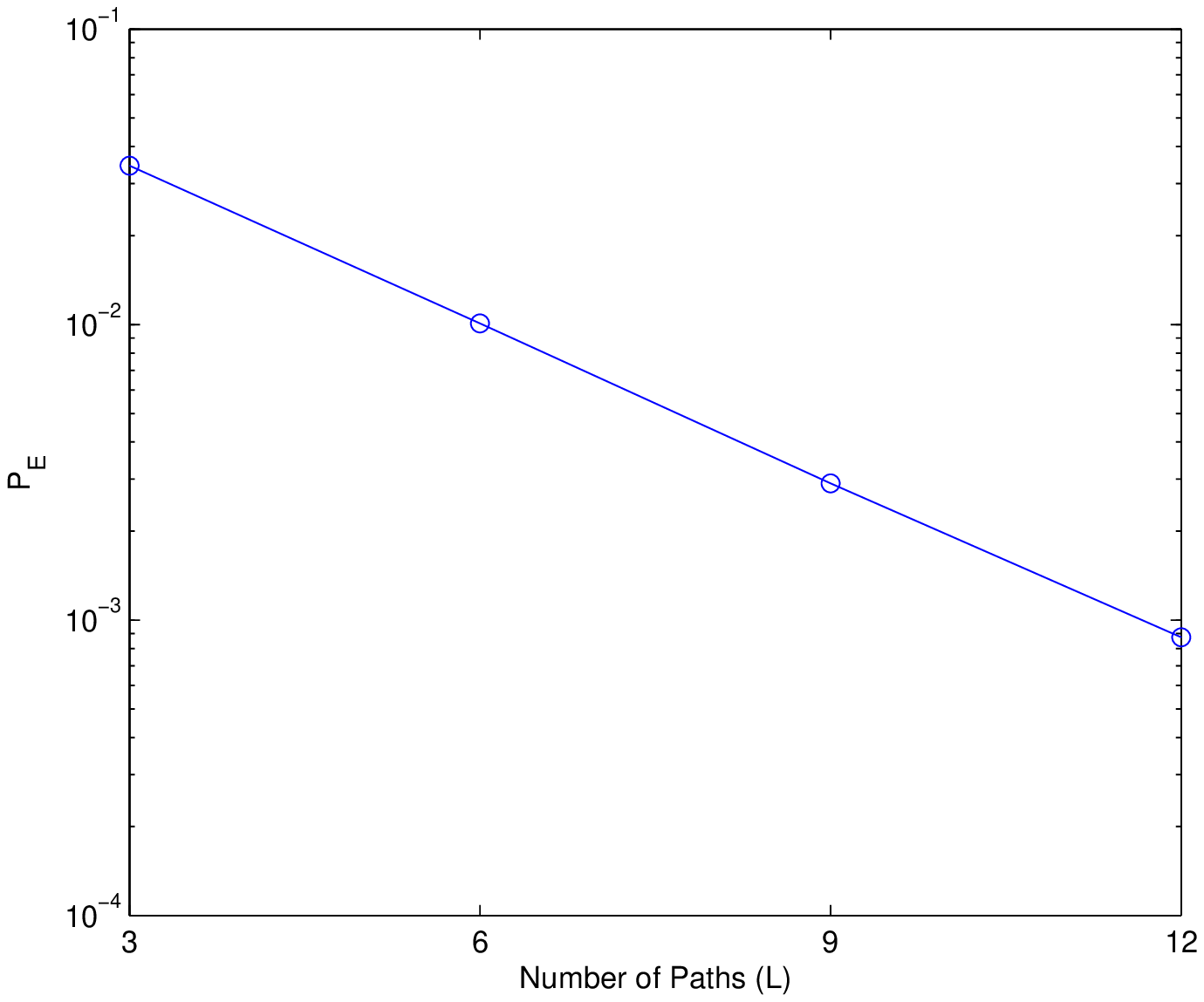}}
     \vspace{0.0in}
     \subfigure[]{
          \label{fig:EttaMax}
          \includegraphics[width=0.45\textwidth]{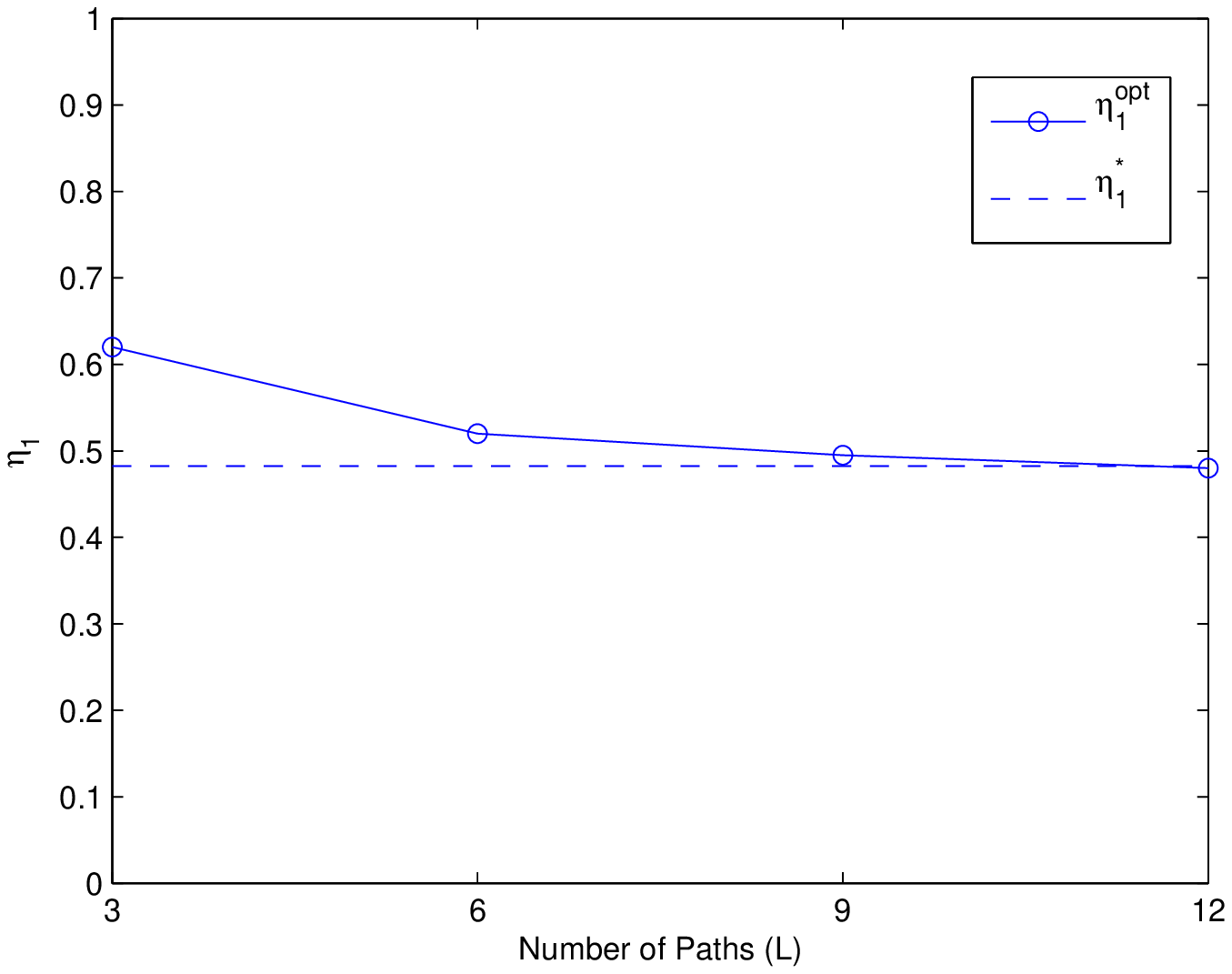}}
     \caption{(a) $P_{E}$ versus $L$ for the combination of two path types, one third from type
I and the rest from type II. (b) The normalized aggregated weight of type I paths in the optimal rate
allocation ($\eta_{1}^{opt}$), compared with the value of $\eta_1$ which maximizes the exponent of equation~\eqref{equation:MaxEttaMinBeta} ($\eta_{1}^{\star}$). }
     \label{fig:TwoTypesOptimization}
\end{figure}

\textbf{Remark III.} Theorem II can be interpreted as follows. For
large values of $L$, adding a new type of path contributes to the
path diversity \textit{iff} the path satisfies the quality
constraint $\alpha > \mathbb E \{ x \}$, where $x$ is the percentage
of time that the path spends in the bad state in the time interval
$[0, T]$. Only in this case, adding the new type of path
exponentially improves the performance of the system in terms of the
probability of irrecoverable loss.

\textbf{Remark IV.} Observing the exponent coefficient corresponding
to the optimum allocation vector $\boldsymbol{\eta}^\star$, we can see that the
typical error event occurs when the ratio of the lost packets on all
types of paths is the same as the total fraction of the lost
packets, $\alpha$. However, this is not the case for any arbitrary
rate allocation vector $\boldsymbol\eta$.

\begin{figure}
    \centering
    \includegraphics[scale=0.45]{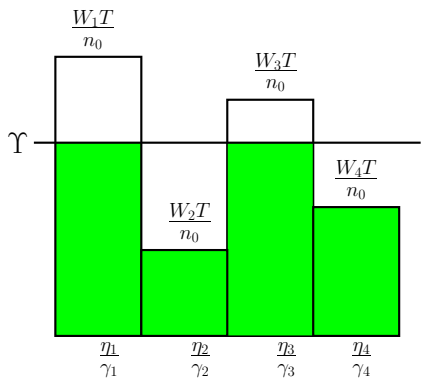}
    \caption{WaterFilling algorithm over identical paths with four different bandwidth constraints.}
    \label{fig:WaterFilling}
\end{figure}

\textbf{Remark V.} An interesting extension of Theorem II is the case where 
all types have identical erasure patterns ($u_j(x)=u_k(x)$ for $\forall~1 \leq j,k \leq J$ and $\forall x$), but 
different bandwidth constraints. Adopting the notation of Theorem II, the bandwidth constraint on $\eta_j$ can be 
written as $\frac{\eta_j n_0 L}{T \gamma_j L} \leq W_j$, where $W_j$ is the maximum bandwidth for a path of 
type $j$. Let us define $\boldsymbol{\tilde{\eta}}^{\star}$ as the allocation vector which 
maximizes the objective function of Theorem II ($g(\boldsymbol{\eta})$), and satisfies the bandwidth constraints 
too. $\boldsymbol{\eta}^{\star}$ is also defined as the maximizing vector for the unconstrained problem in 
Theorem II. According to equation~\eqref{equation:etaStar}, we have $\eta^{\star}_j=\gamma_j$ for 
$\forall 1 \leq j \leq J$. It is obvious 
that $\boldsymbol{\tilde{\eta}}^{\star}=\boldsymbol{\eta}^{\star}$ if $\eta^{\star}_j \leq \frac{\gamma_j W_j T}{n_0}$ for 
all $j$. In case $\eta^{\star}_j$ does not satisfy the bandwidth constraint for 
some $j$, $\boldsymbol{\tilde{\eta}}^{\star}$ can be found by the water-filling algorithm. More accurately, we have 
\begin{equation}
\tilde{\eta}^{\star}_j=\left\{ \begin{array}{ll}
                               \dfrac{\gamma_j W_j T}{n_0} & \mbox{if }\tilde{\eta}^{\star}_j \leq \gamma_j \Upsilon \\
                               \gamma_j \Upsilon & \mbox{if }\tilde{\eta}^{\star}_j < \dfrac{\gamma_j W_j T}{n_0}
                               \end{array} \right.
\label{equation:WaterFilling}
\end{equation}
where $\Upsilon$ can be found by imposing the 
condition $\sum_{j=1}^{J} \tilde{\eta}^{\star}_j=1$. Figure~\ref{fig:WaterFilling} depicts water-filling among identical 
paths with four different bandwidth constraints. Proof of equation~\eqref{equation:WaterFilling} can be found in appendix~\ref{section:ProofRemarkV}.

Figure~\ref{fig:TwoTypes} shows $P_E$ of the optimum rate allocation
versus $L$ for a system consisting of two types of path. The optimal
rate allocation is found by exhaustive search among all possible
allocation vectors. The block transmission time is $T=200ms$. The
block length is proportional to the number of paths as $N=20L$. The
average good burst, $\mu_g$, is selected such that we have $\mu_g
T=\frac{1}{5}$ for both types of paths. $\gamma_1=\frac{1}{3}$ of
the paths (of the first type) benefit from shorter bad bursts and
lower error probability of $\pi_{b,1}=0.015$, and the rest (the
second type) suffer from longer congestion bursts resulting in a
higher error probability of $\pi_{b,2}=0.025$. The coding overhead
is $\alpha=0.1$. The figure depicts a linear behavior in
semi-logarithmic scale with the exponent of $0.403$, which is
comparable to $0.389$ resulted from~\eqref{equation:etaStar}.


In the scenario of Fig.~\ref{fig:TwoTypes}, let us denote
$\eta_{1}^{\star}$ as the value of of the first element of
$\boldsymbol \eta$ in equation~\eqref{equation:etaStar}. Obviously,
$\eta_{1}^{\star}$ does not depend on $L$. Moreover,
$\eta_{1}^{opt}$ is defined as the normalized aggregated weight of
type I paths in the optimal rate allocation.
Figure~\ref{fig:EttaMax} compares $\eta_{1}^{opt}$ with
$\eta_{1}^{\star}$ for different number of paths. It is observed
that $\eta_{1}^{opt}$ converges rapidly to $\eta_1^\star$ as $L$
grows. Figure~\ref{fig:TwoTypes} also verifies that the allocation
vector candidate $\eta^\star$ proposed by Theorem II indeed meets
the optimal allocation vector for large values of $L$.

\section{Suboptimal Rate Allocation}
\label{section:SubOptimalRateAlloc}

In order to compute the complexity of the rate allocation problem,
we focus our attention on the original discrete formulation in
subsection~\ref{subsection:RateAllocProb}. According to the model of
subsection~\ref{subsection:NonIdenticalPaths}, we assume the
available paths are from $J$ types, $L_j$ paths from type $j$, such
that $\sum_{j=1}^{J}L_j=L$. Obviously, all the paths from the same
type should have equal rate. Therefore, the rate allocation problem
is turned into finding the vector $\mathbf{N}=(N_1,\dots,N_J)$ such
that $\sum_{j=1}^{J}N_j=N$, and $0 \leq N_j \leq L_j W_j T$ for all
$j$. $N_j$ denotes the number of packets assigned to all the paths
of type $j$. Let us temporarily assume that all paths have enough
bandwidth such that $N_j$ can vary from $0$ to $N$ for all $j$.
There are $\binom{N+J-1}{J-1}$ $L$-dimensional non-negative vectors
of the form $(N_1,\dots,N_J)$ which satisfy the equation
$\sum_{j=1}^{J}N_j=N$ each representing a distinct rate allocation.
Hence, the number of candidates is exponential in terms of $J$.

First, we prove the problem of rate allocation is NP~\cite{Papadimitriou1994} in 
the sense that $P_E$ can be computed in
polynomial time for any candidate vector
$\mathbf{N}=(N_1,\dots,N_J)$. Let us define $P_e^{\mathbf{N}}(k,j)$
as the probability of having more than $k$ errors over the paths of
types $1$ to $j$ for a specific allocation vector $\mathbf{N}$. We
also define $Q_j(n,k)$ as the probability of having exactly $k$
errors out of the $n$ packets sent over the paths of type $j$.
$Q_j(n,k)$ can be computed and stored for all path types and values
of $n$ and $k$ with polynomial complexity as explained in
appendices~\ref{section:DiscAnalOnePath}
and~\ref{section:DiscAnalOneType}. Then, the following recursive
formula holds for $P_e^{\mathbf{N}}(k,j)$
\begin{eqnarray}
P_e^{\mathbf{N}}(k,j) & = &  \hspace{-0.3cm} \left\{ \begin{array}{lr}
                            \displaystyle\sum_{i=0}^{N_j}{Q_j(N_j,i)P_e^{\mathbf{N}}(k-i,j-1)} & \mbox{if $k\geq 0$} \\
                            1 & \mbox{if $k<0$}\end{array} \right.
\nonumber \\
P_e^{\mathbf{N}}(k,1) & = & \hspace{-0.3cm} \sum_{i=k+1}^{N_1}{Q_1(N_1,i)} .
\label{equation:PeNkj}
\end{eqnarray}
To compute $P_e^{\mathbf{N}}(K,J)$ by the above recursive formula,
we apply a well-known technique in the theory of algorithms called
\textit{memoization}~\cite{Cormen2001}. Memoization works by storing
the computed values of a recursive function in an array. By keeping
this array in the memory, memoization avoids recomputing the
function for the same arguments when it is called later. To compute
$P_e^{\mathbf{N}}(K,J)$, an array of size $O(KJ)$ is required. This
array should be filled with the values of $P_e^{\mathbf{N}}(k,j)$
for $0<k\leq K$, and $1 \leq j \leq J$. Computing
$P_e^{\mathbf{N}}(k,j)$ requires $O(K)$ operations assuming the
values of $P_e^{\mathbf{N}}(i,j-1)$ and $Q_j(N_j,i)$ and
$\sum_{i=k+1}^{N_j}Q_j(N_j,i)$ are already computed for $0 \leq i
\leq k$. Thus, $P_e^{\mathbf{N}}(K,J)$ can be computed with the
complexity of $O(K^2 J)$ if the values of $Q_j(N_j,k)$ are given for
all $N_j$ and $0 \leq k \leq K$. Following appendix
\ref{section:DiscAnalOneType}, we note that for each $j$,
$Q_j(N_j,k)$ for $0 \leq k \leq K$ is computed offline with the
complexity of $O(K^2 L_j) + O\left(\frac{N_j}{L_j} K\right)$. Hence,
the total complexity of computing $P_e^{\mathbf{N}}(K,J)$ adds up to
\begin{eqnarray}
& & O(K^2 J)+\sum_{j=1}^{J}O\left(K^2 L_j+\frac{N_j}{L_j} K\right) \nonumber \\
& \stackrel{(a)}{=} & O(K^2J)+\sum_{j=1}^{J}O\left(K^2 L_j+ N_j K\right) \nonumber \\
& \stackrel{(b)}{=} & O\left( K^2 L + KN \right)
\label{equation:complexity}
\end{eqnarray}
where $(a)$ follows from the fact that $\frac{N_j}{L_j}<N_j$, and
the term $O(K^2J)$ is omitted in $(b)$ since we know that $J<L$.

Now, we propose a suboptimal polynomial time algorithm to estimate
the best path allocation vector, $\mathbf{N}^{opt}$. Let us define
$P_e^{opt}(n,k,j)$ as the probability of having more than $k$ errors
for a block of length $n$ over the paths of types $1$ to $j$
minimized over all possible rate allocations
($\mathbf{N}=\mathbf{N}^{opt}$). First, we find a lowerbound
$\hat{P}_e(n,k,j)$ for $P_e^{opt}(n,k,j)$ from the following
recursive formula
\begin{eqnarray}
\hat{P}_e(n,k,j) & \hspace{-0.2cm} = & \hspace{-0.2cm} \left\{ \begin{array}{lr}
                             \displaystyle \min_{0 \leq n_j \leq \min{\left \{n, \lfloor  L_j W_j T\rfloor \right\}}} \displaystyle \sum_{i=0}^{n_j}Q_j(n_j,i)\cdot & \\
                             ~~~~~\hat{P}_e(n-n_j,k-i,j-1) & \hspace{-0.9cm} \mbox{if $k > 0$} \\
                              & \\
                             1 & \hspace{-0.9cm} \mbox{if $k \leq 0$}\end{array} \right.
\nonumber \\
\hat{P}_e(n,k,1) & \hspace{-0.2cm} = & \hspace{-0.2cm} \sum_{i=k+1}^{n}{Q_1(n,i)} .
\label{equation:PeHatnkj}
\end{eqnarray}
Using memoization technique, we need an array of size $O(NKJ)$ to
store the values of $\hat{P}_e(n,k,j)$ for $0 < n \leq N$, $0<k\leq
K$, and $1 \leq j \leq J$. According to the recursive definition
above, computing $\hat{P}_e(n,k,j)$ requires $O(NK)$ operations
assuming the values of $Q_j(n_j,i)$ and $\hat{P}_e(n-n_j,k-i,j-1)$
and $\sum_{i=k+1}^{n_j}Q_j(n_j,i)$ are already computed for all $i$
and $n_j$. Thus, it is easy to verify that $\hat{P}_e(N,K,J)$ can be
computed with the complexity of $O(N^2K^2J)$ when the values of
$Q_j(n_j,i)$ are given for all $0<n_j \leq n$ and $0 \leq i \leq K$.
According to appendix \ref{section:DiscAnalOneType}, for each $1
\leq j \leq J$, and for each $0<n_j \leq N$, $Q_j(n_j,i)$ for all $0
\leq i \leq n_j$ is computed offline with the complexity of $O(n_j^2
L_j) + O\left(\frac{n_j}{L_j} n_j\right)=O(n_j^2 L_j)$. Thus,
computing $Q_j(n_j,i)$ for all $1 \leq j \leq J$, and $0<n_j \leq
N$, and $0 \leq i \leq n_j$, has the complexity of $\sum_{j=1}^{J}
\sum_{n_j=1}^{N} O(n_j^2 L_j)= O(N^3 L)$. Finally,
$\hat{P}_e(N,K,J)$ can be computed with the total complexity of
$O(N^2K^2J+N^3 L)$.

The following lemma guarantees that $\hat{P}_e(n,k,j)$ is in fact a
lowerbound for $P_e^{opt}(n,k,j)$.

\textbf{Lemma V.} $P_e^{opt}(n,k,j) \geq \hat{P}_e(n,k,j)$. The
proof is given in appendix~\ref{section:ProofLemmaV}.

The following algorithm recursively finds a suboptimum allocation
vector $\mathbf{\hat{N}}$ based on the lowerbound of Lemma V.
\textit{\begin{enumerate}
\item[(1):] Initialize $j \leftarrow J$, $n \leftarrow N$, $k \leftarrow K$.
\item[(2):] Set
\begin{eqnarray}
\hat{N}_j & = & \argmin_{0 \leq n_j \leq \min{\left \{ n, \lfloor L_j W_j T \rfloor \right \}}} \sum_{i=0}^{n_j} Q_j(n_j,i)\cdot \nonumber \\
    &   & \hat{P}_e(n-n_j,k-i,j-1) \nonumber \\
K_j & = & \argmax_{0 \leq i \leq \hat{N}_j}{Q_j(\hat{N}_j,i) \hat{P}_e(n-\hat{N}_j,k-i,j-1)} \nonumber
\end{eqnarray}
\item[(3):] Update $n \leftarrow n-\hat{N}_j$, $k \leftarrow k-K_j$, $j \leftarrow j-1$.
\item[(4):] If $j>1$ and $k \geq 0$, goto (2).
\item[(5):] For $m=1$ to $j$, set $\hat{N}_m \leftarrow \lfloor \dfrac{n}{j} \rfloor $.
\item[(6):] $\hat{N}_j \leftarrow \hat{N}_j+\mbox{Rem}(n,j)$ where $\mbox{Rem}(a,b)$ denotes the remainder of dividing $a$ by $b$.
\end{enumerate}}

Intuitively speaking, the above algorithm tries to recursively find
the typical error event ($K_j$'s) which has the maximum contribution
to the error probability, and assigns the rate allocations ($\hat
N_j$'s) such that the estimated typical error probability ($\hat
P_e$) is minimized. Indeed, Lemma V shows that the estimate used in
the algorithm ($\hat P_e$) is a lower-bound for the minimum
achievable error probability ($P_e^{opt}$). Comparing
(\ref{equation:PeHatnkj}) and the step (2) of our algorithm, we
observe that the values of $\hat N_j$ and $K_j$ can be found in
$O(1)$ during the computation of $\hat P_e(N, K, J)$. Hence,
complexity of the proposed algorithm is the same as that of
computing $\hat P_e(N, K, J)$, $O(N^2K^2J+N^3 L)$.

The following theorem guarantees that the output of the above
algorithm converges to the asymptotically optimal rate allocation
introduced in Theorem II of
section~\ref{subsection:NonIdenticalPaths}, and accordingly, it
performs optimally for large number of paths.

\textbf{Theorem III.} Consider a point-to-point connection over the
network with $L$ independent paths from the source to the
destination, each modeled as a Gilbert-Elliot cell with a large enough bandwidth
constraint. The paths are from $J$ different types, $L_j$ paths from
the type $j$. Assume a block FEC of the size $[N, K]$ is sent during
an interval time $T$.  For fixed values of
$\gamma_j=\frac{L_j}{L},~n_0=\frac{N}{L},~k_0=\frac{K}{L},~T$ and
asymptotically large number of paths ($L$) we have
\begin{enumerate}
\item $\hat{P}_e(N,K,J)\doteq P_e^{opt}(N,K,J) \doteq e^{ -L \sum_{j=1}^{J} \gamma_j u_j\left( \alpha \right) }$
\item $\dfrac {\hat{N}_j} {N} = \eta_j^\star+o(1)$
\item $\frac{K_j}{\hat{N}_j}=\alpha + o(1) $ for $\alpha > \mathbb{E}\{x_j\}$.
\end{enumerate}
where $\alpha=\frac{k_0}{n_0}$ and $u_j()$ are defined
in subsections~\ref{subsection:IdenticalPaths}
and~\ref{subsection:NonIdenticalPaths}. $\hat{P}_e(N,K,J)$ is the
lowerbound for $P_e^{opt}(n,k,j)$ defined in equation
\eqref{equation:PeHatnkj}. $\hat{N}_j$ is the total number of
packets assigned to the paths of type $j$ by the suboptimal rate
allocation algorithm. $\eta_j^\star$ is the asymptotically optimal
rate allocation given in equation~\eqref{equation:etaStar}. $K_j$ is
also defined in the step (2) of the algorithm. The notation
$f(L)=o(g(L))$ means $\lim_{L \to \infty} \frac{f(L)}{g(L)} = 0$.
The proof can be found in appendix~\ref{section:ProofTheoremIII}.

\begin{figure}
    \centering
    \includegraphics[scale=0.50]{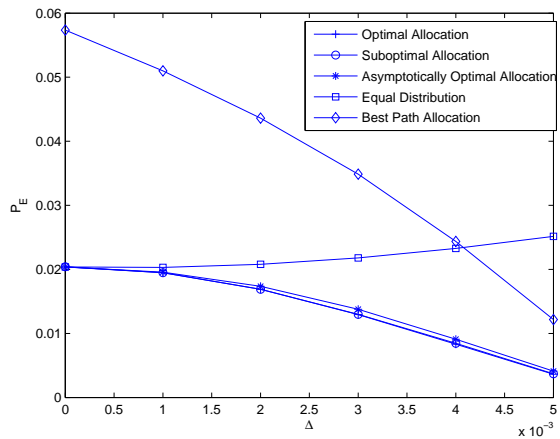}
    \caption{Optimal and suboptimal rate allocations are compared with equal distribution
and best path allocation schemes for different values of $\Delta$}
    \label{fig:SubOptimal}
\end{figure}

The proposed algorithm is compared with four other allocation
schemes over $L=6$ paths in Fig.~\ref{fig:SubOptimal}. The optimal
method uses exhaustive search over all possible allocations.
`\textit{Best Path Allocation}' assigns everything to the best path
only, ignoring the rest. `\textit{Equal Distribution}' scheme
distributes the packets among all paths equally. Finally, the
`\textit{Asymptotically Optimal}' allocation assigns the rates based
on equation~\eqref{equation:etaStar}. The block length and the
number of information packets are assumed to be $N=100$ and $K=90$,
respectively. The overall rate is $S_{req}=1000 pkt/sec$ which
results in $T=100ms$. The average good burst, $\mu_g$, is selected
such that we have $\mu_g T=\frac{1}{5}$. However, quality of the
paths are different as they have different average bad burst
durations. Packet error probability of the paths are listed as
$[0.0175\pm\frac{\Delta}{2},0.0175\pm\frac{3\Delta}{2},0.0175\pm\frac{5\Delta}{2}]$,
such that the median is fixed at $0.0175$. $\Delta$ is also defined
as a measure of deviation from this median. $\Delta=0$ represents
the case where all the paths are identical. The larger is $\Delta$,
the more variety we have among the paths and the more diversity gain
might be achieved using a judicious rate allocation.


As seen, our suboptimal algorithm tracks the optimal algorithm so
closely that the corresponding curves are not easily distinguishable
over a wide range. However, the '\textit{Asymptotically Optimal}'
rate allocation results in lower performance since there is only one
path from each type which makes the asymptotic analysis assumptions
invalid. When $\Delta=0$, `\textit{Equal Distribution}' scheme
obviously coincides with the optimal allocation. This scheme
eventually diverges from the optimal algorithm as $\Delta$ grows.
However, it still outperforms the best path allocation method as
long as $\Delta$ is not too large. For very large values of
$\Delta$, the best path dominates all the other ones, and we can
ignore the rest of the paths. Hence, the best path allocation
eventually converges to the optimal scheme when $\Delta$ increases.

\section{Conclusion}
\label{section:Conclusion}

In this work, we have studied the performance of forward error
correction over a block of packets sent through multiple independent
paths. It is known that \textit{Maximum Distance Separable} (MDS)
block codes are optimum over our \textit{End-to-End Channel} model,
and any other erasure channel with or without memory, in the sense
that their probability of error is minimum among all block codes of
the same size~\cite{Fashandi2008isit,Fashandi20083}. Adopting
MDS codes, the probability of irrecoverable loss, $P_E$, is analyzed
for the cases of a single path, multiple identical, and multiple
non-identical paths based on the discrete to continuous relaxation.
When there are $L$ identical paths, $P_E$ is upperbounded using
large deviation theory. This bound is shown to be exponentially
tight in terms of $L$. The asymptotic analysis shows that the
exponential decay of $P_E$ with $L$ is still valid in the case of
non-identical paths. Furthermore, the optimal rate allocation
problem is solved in the asymptotic case where $L$ is very large. It
is seen that for the optimal rate allocation, each path is assigned
a positive rate \textit{iff} its quality is above certain threshold.
The quality of a path is defined as the percentage of the time it
spends in the bad state. Finally, we focus on the problem of optimum
rate allocation when $L$ is not necessarily large. A heuristic
suboptimal algorithm is proposed which computes a near-optimal
allocation in polynomial time. For large values of $L$, the result
of this algorithm converges to the optimal solution. Moreover,
simulation results are provided which verify the validity of our
theoretical analyses in several practical scenarios, and also show
that the proposed suboptimal algorithm approximates the optimal
allocation very closely.

\appendices
\section{Proof of Lemma I}
\label{section:ProofLemmaI}
\noindent 1) We define the function $v(\lambda)$ as
\begin{equation}
v(\lambda)=\frac  {\mathbb{E}\{xe^{\lambda x}\}} {\mathbb{E}\{e^{\lambda x}\}}.
\end{equation}
Then, the first derivative of $v(\lambda)$ will be
\begin{eqnarray}
\frac{\partial }{\partial \lambda} v(\lambda) & = & \frac
{\mathbb{E}\{x^2e^{\lambda x}\} \mathbb{E}\{e^{\lambda x}\} - [\mathbb{E}\{xe^{\lambda x}\}]^2} {[\mathbb{E}\{e^{\lambda x}\}]^2}.
\label{equation:PartialLambda}
\end{eqnarray}
According to Cauchy-Schwarz inequality, the following statement is
always true for any two functions of $f()$ and $g()$
\begin{equation}
\left(\int_{x}f(x)g(x)dx\right)^2<\int_{x}f^2(x)dx\int_{x}g^2(x)dx
\end{equation}
unless $f(x)=K g(x)$ for a constant $K$ and all values of $x$. If we
choose $f(x)=\sqrt{x^2Q(x)e^{x\lambda}}$ and
$g(x)=\sqrt{Q(x)e^{x\lambda}}$, they can not be proportional to each
other for all values of $x$. Therefore, the numerator of
equation~\eqref{equation:PartialLambda} has to be strictly positive
for all $\lambda$. Since the function $v(\lambda)$ is strictly
increasing, it has an inverse $v^{-1}(\alpha)$ which is also
strictly increasing. Moreover, the non-linear equation
$v(\lambda)=\alpha$ has a unique solution of the form
$\lambda=v^{-1}(\alpha)=l(\alpha)$.

\vspace{\baselineskip}

\noindent 2) To show that $l(\alpha=0)=-\infty$, we prove an
equivalent statement of the form $\lim_{\lambda \to
-\infty}v(\lambda)=0$. Since $x$ is a random variable in the range
$[0,1]$ with the probability density function $Q(x)$, for any
$0<\epsilon<1$, we can write
\begin{eqnarray}
\lim_{\lambda \to -\infty} v(\lambda) & \hspace{-0.3cm} = & \hspace{-0.3cm} \lim_{\lambda \to -\infty}
\dfrac{ \int_{0}^{\epsilon}xQ(x)e^{x\lambda}dx + \int_{\epsilon}^{1}xQ(x)e^{x\lambda}dx }
{ \int_{0}^{1}Q(x)e^{x\lambda}dx } \nonumber \\
& \hspace{-0.3cm} \leq &  \hspace{-0.3cm} \lim_{\lambda \to -\infty}
\dfrac{ \int_{0}^{\epsilon}xQ(x)e^{x\lambda}dx } { \int_{0}^{\epsilon}Q(x)e^{x\lambda}dx } + \dfrac{ \int_{\epsilon}^{1}xQ(x)dx } { \int_{0}^{\epsilon}Q(x)e^{(x-\epsilon)\lambda}dx } \nonumber \\
& \hspace{-0.3cm} \stackrel{(a)}{=} & \hspace{-0.3cm} \lim_{\lambda \to -\infty}
\dfrac{ \int_{0}^{\epsilon}xQ(x)e^{x\lambda}dx } { \int_{0}^{\epsilon}Q(x)e^{x\lambda}dx } \nonumber \\
& \hspace{-0.3cm} \stackrel{(b)}{=} & \hspace{-0.3cm} \lim_{\lambda \to -\infty}
\dfrac{ x_1 Q(x_1)e^{\lambda x_1} } { Q(x_2)e^{\lambda x_2} }
\label{equation:TwoTerms}
\end{eqnarray}
for some $x_1 , x_2 \in [0,\epsilon]$. $(a)$ follows from the fact
that for $x\in[0,\epsilon]$, $(x-\epsilon)\lambda \to +\infty$ when
$\lambda \to -\infty$, and $(b)$ is a result of the mean value
theorem for integration~\cite{Rudin1976}. This theorem states that
for every continuous function $f(x)$ in the interval $[a,b]$, we
have
\begin{eqnarray}
\exists~ x_0 \in [a,b] & s.t. & \int_{a}^{b}f(x)dx=f(x_0)[b-a].
\end{eqnarray}
Equation~\eqref{equation:TwoTerms} is valid for any arbitrary
$0<\epsilon<1$. If we choose $\epsilon\to 0$, $x_1$ and $x_2$ are
both squeezed in the interval $[0,\epsilon]$. Thus, we have
\begin{equation}
\lim_{\lambda \to -\infty} v(\lambda) \leq
\lim_{\lambda \to -\infty} \lim_{\epsilon \to 0} \frac{ x_1 Q(x_1)e^{\lambda x_1} } { Q(x_2)e^{\lambda x_2} }
= \lim_{\epsilon \to 0} x_1 = 0
\label{equation:LimMinusV}
\end{equation}
Based on the distribution of $x$, $v(\lambda)$ is obviously
non-negative for any $\lambda$. Hence, the inequality
in~\eqref{equation:LimMinusV} can be replaced by equality.

\vspace{\baselineskip}

\noindent 3) By observing that $v(\lambda=0)=\mathbb{E}\{x\}$, it is
obvious that $l(\alpha=\mathbb{E}\{x\})=0$.

\vspace{\baselineskip}

\noindent 4) To show that $l(\alpha=1)=+\infty$, we prove the
equivalent statement of the form $\lim_{\lambda \to
+\infty}v(\lambda)=1$. For any $0<\epsilon<1$ and
$x\in[1-\epsilon,1]$, $(x-1+\epsilon)\lambda \to +\infty$ when
$\lambda \to +\infty$. Then, defining $\zeta=1-\epsilon$, we have
\begin{equation}
\lim_{\lambda \to +\infty} \dfrac{ \int_{0}^{\zeta}xQ(x)e^{x\lambda}dx }
{ \int_{0}^{1}Q(x)e^{x\lambda}dx } \leq
\lim_{\lambda \to +\infty} \dfrac{ \int_{0}^{\zeta}xQ(x)dx }
{ \int_{\zeta}^{1}Q(x)e^{(x-\zeta)\lambda}dx } = 0.
\label{equation:LimLemma1}
\end{equation}
Since the fraction in~\eqref{equation:LimLemma1} is obviously
non-negative for all $\lambda$, this inequality can be replaced by
an equality. Similarly, we have
\begin{equation}
\lim_{\lambda \to +\infty} \dfrac{ \int_{0}^{\zeta}Q(x)e^{x\lambda}dx }
{ \int_{\zeta}^{1}xQ(x)e^{x\lambda}dx } \leq
\lim_{\lambda \to +\infty} \dfrac{ \int_{0}^{\zeta}Q(x)dx }
{ \int_{\zeta}^{1}xQ(x)e^{(x-\zeta)\lambda}dx } = 0.
\label{equation:LimLemma2}
\end{equation}
which can also be replaced by equality. Now, the
limit of $v(\lambda)$ is written as
\begin{eqnarray}
\displaystyle\lim_{\lambda \to +\infty} v(\lambda) & \hspace{-0.3cm} = & \hspace{-0.4cm} \displaystyle\lim_{\lambda \to +\infty}
\dfrac{ \int_{0}^{\zeta}xQ(x)e^{x\lambda}dx + \int_{\zeta}^{1}xQ(x)e^{x\lambda}dx }
{ \int_{0}^{1}Q(x)e^{x\lambda}dx } \nonumber \\
& \hspace{-0.3cm} \stackrel{(a)}{=}  & \hspace{-0.4cm} \displaystyle\lim_{\lambda \to +\infty}
\dfrac{ \int_{\zeta}^{1}xQ(x)e^{x\lambda}dx } { \int_{0}^{1}Q(x)e^{x\lambda}dx } \nonumber \\
& \hspace{-0.3cm} \stackrel{(b)}{=}  & \hspace{-0.4cm} \left( \displaystyle\lim_{\lambda \to +\infty}
\dfrac { \int_{0}^{\zeta}Q(x)e^{x\lambda}dx + \int_{\zeta}^{1}Q(x)e^{x\lambda}dx }
{ \int_{\zeta}^{1}xQ(x)e^{x\lambda}dx } \right) ^ {-1} \nonumber \\
& \hspace{-0.3cm} \stackrel{(c)}{=}  & \hspace{-0.4cm} \left( \displaystyle\lim_{\lambda \to +\infty}
\dfrac{ \int_{\zeta}^{1}Q(x)e^{x\lambda}dx }
{ \int_{\zeta}^{1}xQ(x)e^{x\lambda}dx } \right)^{-1} \nonumber \\
& \hspace{-0.3cm} \stackrel{(d)}{=} & \hspace{-0.4cm} \left( \displaystyle\lim_{\lambda \to +\infty}
\dfrac{ Q(x_1)e^{x_1 \lambda} } { x_2Q(x_2)e^{x_2 \lambda} } \right)^{-1}
\label{equation:TwoTerms2}
\end{eqnarray}
for some $x_1 , x_2 \in [1-\epsilon,1]$. $(a)$ follows from
equation~\eqref{equation:LimLemma1}, and $(b)$ is valid since the
final result shows that $\lim_{\lambda \to +\infty} v(\lambda)$ is
finite and non-zero~\cite{Rudin1976}. $(c)$ follows from
equation~\eqref{equation:LimLemma2}, and $(d)$ is a result of the
mean value theorem for integration. If we choose $\epsilon\to 0$,
$x_1$ and $x_2$ are both squeezed in the interval $[1-\epsilon,1]$.
Then, equation~\eqref{equation:TwoTerms2} turns into
\begin{equation}
\lim_{\lambda \to +\infty} v(\lambda) = \hspace{-0.1cm}
\left( \lim_{\lambda \to +\infty} \lim_{\epsilon \to 0} \dfrac{ Q(x_1)e^{x_1 \lambda} } { x_2Q(x_2)e^{x_2 \lambda} } \right)^{-1} \hspace{-0.4cm}
= \left( \lim_{\epsilon \to 0} \dfrac{1}{x_2} \right)^{-1} \hspace{-0.4cm} = 1. \nonumber
\label{equation:LimPlusV}
\end{equation}

\vspace{\baselineskip}

\noindent 5) According to equations~\eqref{equation:ChernoffBound}
and~\eqref{equation:AlphaLambda}, the first derivative of
$u(\alpha)$ is
\begin{equation}
\frac{\partial u(\alpha)}{\partial \alpha}=
l(\alpha)+\alpha\frac{\partial l(\alpha)}{\partial \alpha}-
\dfrac{\mathbb{E}\{xe^{\lambda x}\}}{\mathbb{E}\{e^{\lambda x}\}}\frac{\partial l(\alpha)}{\partial \alpha}=
l(\alpha). \nonumber
\end{equation}
\section{Proof of Lemma II}
\label{section:ProofLemmaII}
Based on the definition of probability density function, we have
\begin{eqnarray}
& & \lim_{L\to\infty}-\dfrac{1}{L} \log\left(f_y(\alpha)\right) \nonumber \\
&=& \lim_{L\to\infty} -\dfrac{1}{L} \log\left( \displaystyle\lim_{\delta\to\ 0} \dfrac{\mathbb{P}\{y>\alpha\}-\mathbb{P}\{y>\alpha+\delta\}}{\delta} \right) \nonumber \\
& \stackrel{(a)}{=} & \displaystyle\lim_{\delta\to\ 0} \lim_{L\to\infty} -\dfrac{1}{L} \log\left( \dfrac{\mathbb{P}\{y>\alpha\}-\mathbb{P}\{y>\alpha+\delta\}}{\delta} \right) \nonumber \\
& \geq & \lim_{\delta\to\ 0} \lim_{L\to\infty} \dfrac{1}{L}  \left( -\log\left( \mathbb{P}\{y>\alpha\} \right) + \log\delta \right) \nonumber \\
& \stackrel{(b)}{=} & u(\alpha)
\label{equation:GEQuAlpha}
\end{eqnarray}
where $(a)$ is valid since $\log$ is a continuous function, and both
limitations do exist and are interchangeable. $(b)$ follows from
equation~\eqref{equation:dotEQmain}. The exponent of $f_y(\alpha)$
can be upper-bounded as
\begin{eqnarray}
& \hspace{-0.3cm} & \hspace{-0.3cm} \lim_{L\to\infty}-\dfrac{1}{L} \log\left(f_y(\alpha)\right) \nonumber \\
& \hspace{-0.3cm} \stackrel{(a)}{=} & \hspace{-0.3cm} \lim_{\delta\to\ 0} \lim_{L\to\infty}
\dfrac{ -\log\left( \mathbb{P}\{y>\alpha\} - \mathbb{P}\{y>\alpha+\delta\} \right) + \log\delta }
{ L } \nonumber \\
& \hspace{-0.3cm} \stackrel{(b)}{\leq} & \hspace{-0.3cm} \lim_{\delta\to\ 0} \lim_{L\to\infty}
\dfrac{ -\log\left( e^{-L(u(\alpha)+\epsilon)} - e^{-L(u(\alpha+\delta)-\epsilon)} \right) + \log\delta }
{ L } \nonumber \\
& \hspace{-0.3cm} = & \hspace{-0.3cm} \lim_{\delta\to\ 0} \lim_{L\to\infty}
u(\alpha)+\epsilon-\dfrac {\log\left(1-e^{-L\chi}\right)} {L} \nonumber \\
& \hspace{-0.3cm} \stackrel{(c)}{=} & \hspace{-0.3cm} u(\alpha)+\epsilon
\label{equation:LEQuAlpha}
\end{eqnarray}
where $\chi=u(\alpha+\delta)-u(\alpha)-2\epsilon$. Since $u(\alpha)$
is a strictly increasing function (Lemma I), we can make $\chi$
positive by choosing $\epsilon$ small enough. $(a)$ is valid since
$\log$ is a continuous function, and both limits do exist and are
interchangeable. $(b)$ follows from the definition of limit if $L$
is sufficiently large, and $(c)$ is a result of $\chi$ being
positive. Selecting $\epsilon$ arbitrarily small,
results~\eqref{equation:GEQuAlpha} and~\eqref{equation:LEQuAlpha}
prove the lemma.
\section{Proof of Lemma III}
\label{section:ProofLemmaIII}
According to the definition of infimum, we have
\begin{eqnarray}
& & \lim_{L\to\infty}-\dfrac{\log(H(L))}{L} \nonumber \\
& \geq & \lim_{L\to\infty} -\dfrac {1}{L}
\log\left( e^{-L \displaystyle\inf_{\mathcal{S}}{h(\mathbf{x})} }  \int_{\mathcal{S}}{d\mathbf{x}} \right)
\nonumber \\
& \stackrel{(a)}{=} & \inf_{\mathcal{S}}{h(\mathbf{x})}.
\label{equation:GEQhxSTAR}
\end{eqnarray}
where $(a)$ follows from the fact that $\mathcal S$ is a bounded
region. Since $h(\mathbf{x})$ is a continuous function, it has a
minimum in the bounded closed set  $cl(\mathcal{S})$ which is
denoted by $\mathbf{x^{\star}}$. Due to the continuity of
$h(\mathbf{x})$ at $\mathbf{x^{\star}}$, for any $\epsilon>0$, there
is a neighborhood $\mathcal{B}(\epsilon)$ centered at
$\mathbf{x^{\star}}$ such that any
$\mathbf{x}\in\mathcal{B}(\epsilon)$ has the property of
$|h(\mathbf{x})-h(\mathbf{x^{\star}})|<\epsilon$. Moreover, since
$\mathcal S$ is a convex set, we have $\mbox{vol} \left(
\mathcal{B}(\epsilon)\cap\mathcal{S} \right) > 0$ . Now, we can
write
\begin{eqnarray}
& & \lim_{L\to\infty}-\dfrac{\log(H(L))}{L} \nonumber \\
& \leq & \lim_{L\to\infty} -\dfrac {1}{L}
\log\left( \int_{\mathcal{S} \cap \mathcal{B}(\epsilon) }{ e^{-L h(\mathbf{x})} d\mathbf{x}} \right) \nonumber \\
& \leq & \lim_{L\to\infty} -\dfrac {1}{L}
\log\left( e^{-L (h(\mathbf{x^{\star}})+\epsilon) }  \int_{\mathcal{S}\cap\mathcal{B}(\epsilon)}{d\mathbf{x}} \right)
\nonumber \\
& =    & h(\mathbf{x^{\star}})+\epsilon.
\label{equation:LEQhxSTAR}
\end{eqnarray}
Selecting $\epsilon$ to be arbitrarily small,
\eqref{equation:GEQhxSTAR} and~\eqref{equation:LEQhxSTAR} prove the
lemma.
\section{Proof of Lemma IV}
\label{section:ProofLemmaIV} According to Lemma I, $u_j(x)$ is
increasing and convex for $\forall 1\leq j \leq J$. Thus, the
objective function $f(\boldsymbol{\beta})=\sum_{j=1}^{J}{\gamma_j
u_j(\frac {\beta_j} {\eta_j})}$ is also convex, and the region
$\mathcal{S}_T$ is determined by $J$ convex inequality constraints
and one affine equality constraint. Hence, in this case, KKT
conditions are both necessary and sufficient for
optimality~\cite{Boyd2004}. In other words, if there exist constants
$\phi_j$ and $\nu$ such that
\begin{eqnarray}
\frac{\gamma_j}{\eta_j}l_j(\frac{\beta_j^{\star}}{\eta_j})-\phi_j-\nu=0 & \forall 1\leq j \leq J
\label{equation:KKT1} \\
\phi_j \left[ \eta\mathbb{E}\{x_j\} - \beta_j^{\star} \right]=0 & \forall 1\leq j \leq J
\label{equation:KKT2}
\end{eqnarray}
then the point $\boldsymbol{\beta}^{\star}$ is a global minimum.

Now, we prove that either $\beta_j^{\star}=\eta_j\mathbb{E}\{x_j\}$
for all $1\leq j \leq J$, or
$\beta_j^{\star}>\eta_j\mathbb{E}\{x_j\}$ for all $1\leq j \leq J$.
Let us assume the opposite is true, and there are at least two
elements of the vector $\boldsymbol{\beta}^{\star}$, indexed with
$k$ and $m$, which have the values of
$\beta_k^{\star}=\eta_k\mathbb{E}\{x_k\}$ and
$\beta_m^{\star}>\eta_m\mathbb{E}\{x_m\}$, respectively. For any
arbitrary $\epsilon>0$, the vector $\boldsymbol{\beta}^{\star\star}$
can be defined as below
\begin{equation}
\beta_j^{\star\star}=\left\{\begin{array}{ll}
                    \beta_j^{\star}+\epsilon & \mbox{if } j=k \\
                    \beta_j^{\star}-\epsilon & \mbox{if } j=m \\
                    \beta_j^{\star}          & \mbox{otherwise.}
                    \end{array}  \right.
\end{equation}
Then, we have
\begin{eqnarray}
& \hspace{-0.25cm} & \hspace{-0.25cm} \displaystyle\lim_{\epsilon\to 0} \dfrac{f(\boldsymbol{\beta}^{\star\star})-f(\boldsymbol{\beta}^{\star})}{\epsilon} \nonumber \\
&\hspace{-0.25cm}=& \displaystyle\lim_{\epsilon\to 0} \dfrac{1}{\epsilon}
\left\{ \gamma_k u_k\left( \frac{ \beta_k^{\star}+\epsilon }{ \eta_k } \right) + \gamma_m u_m\left( \dfrac{ \beta_m^{\star}-\epsilon }{ \eta_m } \right) \right. \nonumber \\
&\hspace{-0.25cm} & \hspace{-0.25cm} ~~~~~~~~\left. - \gamma_m u_m\left( \frac{ \beta_m^{\star} }{ \eta_k } \right) \right\} \nonumber \\
&\hspace{-0.25cm}\stackrel{(a)}{=}& \hspace{-0.25cm} \displaystyle\lim_{\epsilon\to 0}
\frac{\gamma_k}{\eta_k} l_k \left( \frac{\beta_k^{\star}+\epsilon' } {\eta_k} \right) -
\frac{\gamma_m}{\eta_m} l_m \left( \frac{\beta_m^{\star}+\epsilon'' } {\eta_m} \right) \nonumber \\
&\hspace{-0.25cm}=& \hspace{-0.25cm} - \frac{\gamma_m}{\eta_m} l_m \left( \frac{\beta_m^{\star}} {\eta_m} \right) < 0
\end{eqnarray}
where $\epsilon',\epsilon''\in [0,\epsilon]$, and $(a)$ follows from
the Taylor's theorem. Thus, moving from $\boldsymbol{\beta}^{\star}$
to $\boldsymbol{\beta}^{\star\star}$ decreases the function which
contradicts the assumption of $\boldsymbol{\beta}^{\star}$ being the
global minimum.

Out of the remaining possibilities, the case where $\beta_j^{\star}=\eta_j\mathbb{E}\{x_j\}$
($\forall 1\leq j \leq J$) obviously agrees with Lemma IV for the special case of
$\nu=0$. Therefore, the lemma can be proved assuming $\beta_j^{\star}>\eta_j\mathbb{E}\{x_j\}$
($\forall 1\leq j \leq J$). Then, equation~\eqref{equation:KKT2} turns into $\phi_j=0$
($\forall 1\leq j \leq J$). By rearranging equation~\eqref{equation:KKT1} and using the
condition $\sum_{j=1}^{J}{\beta_j}=\alpha$, Lemma IV is proved.
\section{Proof of Theorem II}
\label{section:ProofTheoremII} \textbf{Sketch of the proof:} First,
it is proved that $\eta^{\star}_j>0$ if $\mathbb{E}\{x_j\} <
\alpha$. At the second step, we prove that $\eta^{\star}_j=0$, if
$\mathbb{E}\{x_j\} \geq \alpha$. Then, KKT
conditions~\cite{Boyd2004} are applied for the indices $ 1 \leq k
\leq J$ where $\mathbb{E} \{ x_k \} < \alpha$ to find the maximizing
allocation vector, $\boldsymbol{\eta}^\star$.

\textbf{Proof:} The parameter $\nu$ is obviously a function of the
vector $\boldsymbol{\eta}$. Differentiating
equation~\eqref{equation:NuDef} with respect to $\eta_k$ results in
\begin{equation}
\dfrac{ \partial \nu }{ \partial \eta_k } = -
\dfrac{
v_k \left( \dfrac{\nu \eta_k}{\gamma_k} \right)
+
\dfrac{\nu\eta_k}{\gamma_k} v'_k \left( \dfrac{\nu \eta_k}{\gamma_k} \right)
}
{\displaystyle\sum_{j=1}^{J} \dfrac{\eta_j^2}{\gamma_j} v'_j \left( \dfrac{\nu\eta_j}{\gamma_j} \right) }
\label{equation:dNudEtak}
\end{equation}
where $v_j(x)=l^{-1}_j(x)$, and $v'_j(x)$ denotes its derivative
with respect to its argument. The objective function can be
simplified as
\begin{equation}
g(\boldsymbol{{\eta}})=\sum_{j=1}^{J}{\gamma_j u_j(\frac {\beta_j^{\star}} {\eta_j})}=
\sum_{j=1}^{J} { \gamma_j u_j \left( v_j ( \frac {\nu\eta_j} {\gamma_j} ) \right) }.
\end{equation}
$\nu^{\star}$ is defined as the value of $\nu$ corresponding to
$\boldsymbol{\eta}^{\star}$. Next, we show that $\nu^{\star}>0$. Let
us assume the opposite is true, i.e., $\nu^{\star}\leq 0$. Then,
according to Lemma I, we have $v_j ( \frac {\nu^\star \eta_j}
{\gamma_j} ) \leq \mathbb{E}\{x_j\}$ for all $j$ which results in
$g(\boldsymbol{\eta^{\star}})=0$. However, it is possible to achieve
a positive value of $g(\boldsymbol{\eta})$ by setting $\eta_j=1$ for
the one vector which has the property of $\mathbb{E}\{x_j\} <
\alpha$, and setting $\eta_j=0$ for the rest. Thus,
$\boldsymbol{\eta}^{\star}$ can not be the maximal point. This
contradiction proves the fact that $\nu^{\star}>0$.

At the first step, we prove that $\eta^{\star}_j>0$ if
$\mathbb{E}\{x_j\} < \alpha$. Assume the opposite is true for an
index $1\leq k \leq J$. Since $\sum_{j=1}^{J}\eta^{\star}_j=1$,
there should be at least one index $m$ such that $\eta^{\star}_m>0$.
For any arbitrary $\epsilon>0$, the vector
$\boldsymbol{\eta}^{\star\star}$ can be defined as below
\begin{equation}
\eta_j^{\star\star}=\left\{\begin{array}{ll}
                    \epsilon & \mbox{if } j=k \\
                    \eta_j^{\star}-\epsilon & \mbox{if } j=m \\
                    \eta_j^{\star}          & \mbox{otherwise.}
                    \end{array}  \right.
\end{equation}
$\nu^{\star\star}$ is defined as the corresponding value of $\nu$
for the vector $\boldsymbol{\eta}^{\star\star}$. Based on
equation~\eqref{equation:dNudEtak}, we can write
\begin{eqnarray}
& \hspace{-0.25cm} & \Delta\nu = \nonumber \\
& \hspace{-0.25cm} & \nu^{\star\star}-\nu^{\star} =
\label{equation:DeltaNu} \\
& \hspace{-0.25cm} & \dfrac{ v_m \left( \dfrac{\nu^{\star} \eta_m^{\star}}{\gamma_m} \right) +
\dfrac{\nu^{\star} \eta_m^{\star}}{\gamma_m} v'_m \left( \dfrac{\nu^{\star} \eta_m^{\star}}{\gamma_m} \right) - \mathbb{E}\{x_k\} } { \displaystyle\sum_{j=1}^{J} \dfrac{\eta_j^{\star2}}{\gamma_j} v'_j \left( \dfrac{\nu^{\star} \eta_j^{\star}}{\gamma_j} \right) } \epsilon + O(\epsilon^2). \nonumber
\end{eqnarray}
Then, we have
\begin{eqnarray}
& & \hspace{-0.25cm} \displaystyle\lim_{\epsilon\to 0} \dfrac{g(\boldsymbol{\eta}^{\star\star})-g(\boldsymbol{\eta}^{\star})}{\epsilon} \nonumber \\
&=& \hspace{-0.25cm} \displaystyle\lim_{\epsilon\to 0} \dfrac{1}{\epsilon} \left\{
\dfrac{\nu^{\star2}\eta_k^{\star}}{\gamma_k} v'_k \left( \dfrac{\nu^{\star}\eta_k^{\star}}{\gamma_k} \right) \epsilon - \dfrac{\nu^{\star2}\eta_m^{\star}}{\gamma_m} v'_m \left( \dfrac{\nu^{\star}\eta_m^{\star}}{\gamma_m} \right) \epsilon \right. \nonumber \\
& & \hspace{-0.25cm}~~~~~~~~ \left. + ~ \nu^{\star} \Delta\nu \sum_{j=1}^{J}\dfrac{\eta_j^{\star2}}{\gamma_j}
v'_j \left( \dfrac{\nu^{\star}\eta_j^{\star}}{\gamma_j} \right) + O(\epsilon^2)    \right\} \nonumber \\
&\stackrel{(a)}{=}& \hspace{-0.25cm} \nu^{\star} \left\{ v_m \left( \dfrac{\nu^{\star} \eta_m^{\star}}{\gamma_m} \right) -  \mathbb{E}\{x_k\} \right\}
\label{equation:dgdeps}
\end{eqnarray}
where $(a)$ follows from~\eqref{equation:DeltaNu}. If the value
of~\eqref{equation:dgdeps} is positive for an index $m$, moving in
that direction increases the objective function which contradicts
with the assumption of $\boldsymbol{\eta^{\star}}$ being a maximal
point. If the value of~\eqref{equation:dgdeps} is non-positive for
all indexes $m$ whose $\eta_m^{\star}>0$, we can write
\begin{equation}
\mathbb{E}\{x_k\} \geq \sum_{m=1}^{J} \eta_m^{\star}
v_m \left( \dfrac{\nu^{\star} \eta_m^{\star}}{\gamma_m} \right) = \alpha
\end{equation}
which obviously contradicts the assumption of $\mathbb{E}\{x_k\}<\alpha$.

At the second step, we prove that $\eta^{\star}_j=0$ if
$\mathbb{E}\{x_j\} \geq \alpha$. Assume the opposite is true for an
index $1\leq r \leq J$. Since $\sum_{j=1}^{J}\eta^{\star}_j=1$, we
should have $\eta^{\star}_s<1$ for all other indices $s$. For any
arbitrary $\epsilon>0$, the vector
$\boldsymbol{\eta}^{\star\star\star}$ can be defined as
\begin{equation}
\eta_j^{\star\star\star}=\left\{\begin{array}{ll}
                    \eta_j^{\star}-\epsilon & \mbox{if } j=r \\
                    \eta_j^{\star}+\epsilon & \mbox{if } j=s \\
                    \eta_j^{\star}          & \mbox{otherwise.}
                    \end{array}  \right.
\end{equation}
$\nu^{\star\star\star}$ is defined as the corresponding value of
$\nu$ for the vector $\boldsymbol{\eta}^{\star\star\star}$. Based on
equation~\eqref{equation:dNudEtak}, we can write
\begin{eqnarray}
\Delta\nu & \hspace{-0.3cm} = & \hspace{-0.3cm} \nu^{\star\star\star}-\nu^{\star} \nonumber \\
& \hspace{-0.3cm} = & \hspace{-0.3cm} \dfrac{\epsilon}{ \displaystyle\sum_{j=1}^{J} \dfrac{\eta_j^{\star2}}{\gamma_j} v'_j \left( \dfrac{\nu^{\star} \eta_j^{\star}}{\gamma_j} \right) }
\left\{ v_r \left( \dfrac{\nu^{\star} \eta_r^{\star}}{\gamma_r} \right) +
\dfrac{\nu^{\star} \eta_r^{\star}}{\gamma_r} v'_r \left( \dfrac{\nu^{\star} \eta_r^{\star}}{\gamma_r} \right) \right. \nonumber \\
& \hspace{-0.3cm} & \hspace{-0.3cm} \left. - v_s \left( \dfrac{\nu^{\star} \eta_s^{\star}}{\gamma_s} \right)
- \dfrac{\nu^{\star} \eta_s^{\star}}{\gamma_s} v'_s \left( \dfrac{\nu^{\star} \eta_s^{\star}}{\gamma_s} \right) \right\}
+ O(\epsilon^2).
\label{equation:DeltaNu2}
\end{eqnarray}
Then, we have
\begin{eqnarray}
& & \displaystyle\lim_{\epsilon\to 0} \dfrac{g(\boldsymbol{\eta}^{\star\star\star})-g(\boldsymbol{\eta}^{\star})}{\epsilon} \nonumber \\
&=& \displaystyle\lim_{\epsilon\to 0} \dfrac{1}{\epsilon} \left\{
\dfrac{\nu^{\star2}\eta_s^{\star}}{\gamma_s} v'_s \left( \dfrac{\nu^{\star}\eta_s^{\star}}{\gamma_s} \right) \epsilon - \dfrac{\nu^{\star2}\eta_r^{\star}}{\gamma_r} v'_r \left( \dfrac{\nu^{\star}\eta_r^{\star}}{\gamma_r} \right) \epsilon \right. \nonumber \\
& & \left. + ~ \nu^{\star} \Delta\nu \sum_{j=1}^{J}\dfrac{\eta_j^{\star2}}{\gamma_j}
v'_j \left( \dfrac{\nu^{\star}\eta_j^{\star}}{\gamma_j} \right) + O(\epsilon^2) \right\} \nonumber \\
&\stackrel{(a)}{=}& \nu^{\star} \left\{ v_r \left( \dfrac{\nu^{\star} \eta_r^{\star}}{\gamma_r} \right) -
v_s \left( \dfrac{\nu^{\star} \eta_s^{\star}}{\gamma_s} \right) \right\}
\label{equation:dgdeps2}
\end{eqnarray}
where $(a)$ follows from~\eqref{equation:DeltaNu2}. If the value
of~\eqref{equation:dgdeps2} is positive for an index $s$, moving in
that direction increases the objective function which contradicts
with the assumption of $\boldsymbol{\eta^{\star}}$ being a maximal
point . If the value of~\eqref{equation:dgdeps2} is non-positive for
all indices $s$ whose $\eta_s^{\star}>0$, we can write
\begin{equation}
\mathbb{E}\{x_r\} < v_r \left( \dfrac{ \nu^{\star}\eta_r^{\star} }{\gamma_r} \right) \leq
\sum_{s=1}^{J} \eta_s^{\star} v_s \left( \dfrac{\nu^{\star} \eta_s^{\star}}{\gamma_s} \right) = \alpha
\end{equation}
which obviously contradicts the assumption of $\mathbb{E}\{x_r\}\geq\alpha$.

Now that the boundary points are checked, we can safely use the KKT
conditions~\cite{Boyd2004} for all $ 1 \leq k \leq J$, where
$\mathbb{E} \{ x_k \} < \alpha$, to find the maximizing allocation
vector, $\boldsymbol{\eta}^\star$.
\begin{eqnarray}
\zeta &\hspace{-0.3cm} =& \hspace{-0.3cm} \dfrac{\nu^{\star 2}\eta_k^{\star}}{\gamma_k} v'_k \left( \dfrac{\nu^{\star}\eta_k^{\star}}{\gamma_k} \right) +
\nu^{\star} \displaystyle \sum_{j=1}^{J} \dfrac{\eta_j^{\star2}}{\gamma_j} v'_j \left( \dfrac{\nu^{\star}\eta_j^{\star2}}{\gamma_j} \right)
\dfrac{\partial\nu}{\partial\eta_k}|_{\nu=\nu^{\star}} \nonumber \\
& \hspace{-0.3cm} \stackrel{(a)}{=} & \hspace{-0.3cm} -\nu^{\star} v_k \left( \dfrac{\nu^{\star} \eta_k^{\star}}{\gamma_k} \right)
\label{equation:zeta}
\end{eqnarray}
where $\zeta$ is a constant independent of $k$, and $(a)$ follows
from~\eqref{equation:dNudEtak}. Using the fact that
$\sum_{j=1}^{J}\eta_j=1$ together with
equations~\eqref{equation:NuDef} and~\eqref{equation:zeta} results
in
\begin{eqnarray}
& & \zeta=-\alpha\nu^{\star}  \nonumber \\
& & \nu^{\star}=\displaystyle\sum_{\mathbb{E}\{x_j\}<\alpha}\gamma_j l_j(\alpha).
\label{equation:NuStarSolution}
\end{eqnarray}
Combining equations~\eqref{equation:zeta}
and~\eqref{equation:NuStarSolution} results in
equation~\eqref{equation:etaStar} and
$g(\boldsymbol{\eta}^{\star})=\sum_{j=1}^{J}\gamma_j u_j(\alpha)$.
\section{Proof of Remark V}
\label{section:ProofRemarkV}
Based on the arguments similar to the ones in appendix~\ref{section:ProofTheoremII}, it can be shown 
that $\tilde{\eta}^{\star}_j=0$ \textit{iff} $\mathbb{E}\{x_j\} \geq \alpha$. Since all the types are 
identical here, this means $\tilde{\eta}^{\star}_j>0$ for all $j$. Similar to equation~\eqref{equation:zeta}, applying KKT conditions~\cite{Boyd2004}, gives us 
\begin{equation}
v_j \left( \dfrac{\tilde{\nu}^{\star} \tilde{\eta}_j^{\star}}{\gamma_j} \right)= \left\{ \begin{array}{ll}
-\zeta &  \mbox{if }\tilde{\eta}^{\star}_j < \dfrac{\gamma_j W_j T}{n_0} \\
& \\
-\zeta-\sigma_j & \mbox{if }\tilde{\eta}^{\star}_j = \dfrac{\gamma_j W_j T}{n_0}
                                                                                         \end{array} \right.
\label{equation:KKTwaterfilling}
\end{equation}
where $\sigma_j$'s are non-negative parameters~\cite{Boyd2004}. Putting $\Upsilon=\frac{l_j(-\zeta)}{\tilde{\nu}^{\star}}$ proves 
equation~\eqref{equation:WaterFilling}.
\section{Discrete Analysis of One Path}
\label{section:DiscAnalOnePath} $Q(n,k,l)$ is defined as the
probability of having exactly $k$ errors out of the $n$ packets sent
over the path $l$. Depending on the initial state of the path $l$,
$P_g(n,k,l)$ and $P_b(n,k,l)$ are defined as the probabilities of
having $k$ errors out of the $n$ packets sent over this path when we
start the transmission in the good or in the bad state,
respectively. It is easy to see that
\begin{equation}
Q(n,k,l)=\pi_g P_g(n,k,l) + \pi_b P_b(n,k,l).
\label{equation:Qnk}
\end{equation}
$P_g(n,k,l)$ and $P_b(n,k,l)$ can be computed from the following
recursive equations
\begin{eqnarray}
P_b(n,k,l) \hspace{-0.3cm} & = & \hspace{-0.3cm} \pi_{b|b}P_b(n-1,k-1,l)+\pi_{g|b}P_g(n-1,k-1,l) \nonumber \\
P_g(n,k,l) \hspace{-0.3cm} & = & \hspace{-0.3cm} \pi_{b|g}P_b(n-1,k,l)+\pi_{g|g}P_g(n-1,k,l)
\label{equation:PbPgnk}
\end{eqnarray}
with the initial conditions
\begin{eqnarray}
P_g(n,k,l)=0 & & \mbox{for } k \geq n \nonumber \\
P_b(n,k,l)=0 & & \mbox{for } k>n \nonumber \\
P_g(n,k,l)=0 & & \mbox{for } k<0 \nonumber \\
P_b(n,k,l)=0 & & \mbox{for } k \leq 0
\end{eqnarray}
where $\pi_{s_2|s_1}$ is the probability of the channel being in the
state $s_2\in\{g,b\}$ provided that it has been in the state
$s_1\in\{g,b\}$ when the last packet was transmitted.
$\pi_{s_2|s_1}$ has the following values for different combinations
of $s_1$ and $s_2$~\cite{Bolot1999}
\begin{eqnarray}
\pi_{g|g} & = & \pi_g+\pi_b ~ e^{- \dfrac{\mu_g+\mu_b}{S_l} } \nonumber \\
\pi_{b|g} & = & 1-\pi_{g|g} \nonumber \\
\pi_{b|b} & = & \pi_b+\pi_g ~ e^{- \dfrac{\mu_g+\mu_b}{S_l} }\nonumber \\
\pi_{g|b} & = & 1-\pi_{b|b}
\end{eqnarray}
where $S_l$ denotes the transmission rate on the path $l$, i.e., the
packets are transmitted on the path $l$ every $\frac{1}{S_l}$
seconds.

According to the recursive equations in~\eqref{equation:PbPgnk}, to
compute $P_b(n,k,l)$ and $P_g(n,k,l)$ by memoization technique, the
functions $P_b()$ and $P_g()$ should be calculated at the following
set of points denoted as $\mathcal S (n,k)$
\begin{equation}
\mathcal S (n,k) = \left\{ (n',k')~|~ 0  \leq k' \leq k,~ n'-n+k \leq k' \leq n' \right\}. \nonumber
\end{equation}
Cardinality of the set $\mathcal S (n,k)$ is of the order $|\mathcal
S (n,k)|=O\left(k\left(n-k\right)\right)$. Since three operations
are needed to compute the recursive functions $P_b()$ and $P_g()$ at
each point, $P_b(n,k,l)$ and $P_g(n,k,l)$ are computable with the
complexity of $O\left(k\left(n-k\right)\right)$ which give us
$Q(n,k,l)$ according to equation~\eqref{equation:Qnk}.
\section{Discrete Analysis of One Type}
\label{section:DiscAnalOneType} When there are $n$ packets to be
distributed over $L_j$ identical paths of type $j$, uniform
distribution is obviously the optimum. However, since the integer
$n$ may be indivisible by $L_j$, the $L_j$ dimensional vector
$\mathbf{N}$ is selected as
\begin{equation}
N_l=\left\{ \begin{array}{ll}
            \lfloor \dfrac{n}{L_j} \rfloor + 1 & \mbox{for  } 1 \leq l \leq \mbox{Rem}(n,L_j) \\
            & \\
            \lfloor \dfrac{n}{L_j} \rfloor     & \mbox{for  } \mbox{Rem}(n,L_j) < l \leq L_j
            \end{array} \right.
\end{equation}
where $\mbox{Rem}(a,b)$ denotes the remainder of dividing $a$ by
$b$. $\mathbf{N}$ represents the closest integer vector to a uniform
distribution.

$E^{\mathbf{N}}(k,l)$ is defined as the probability of having
exactly $k$ erasures among the $n$ packets transmitted over the
identical paths $1$ to $l$ with the allocation vector $\mathbf{N}$.
According to the definitions of $Q_j(n,k)$ and
$E^{\mathbf{N}}(k,l)$, it is obvious that
$Q_j(n,k)=E^{\mathbf{N}}(k,L_j)$. $E^{\mathbf{N}}(k,l)$ can be
computed recursively as
\begin{eqnarray}
E^{\mathbf{N}}(k,l)& = & \sum_{i=0}^{k}E^{\mathbf{N}}(k-i,l-1)Q(N_l,i,l) \nonumber \\
E^{\mathbf{N}}(k,1)& = & Q(N_1,k,1)
\label{equation:ENkl}
\end{eqnarray}
where $Q(N_l,i,l)$ is given in
appendix~\ref{section:DiscAnalOnePath}. Since all the paths are
assumed to be identical here, $Q(N_l,k,l)$ is the same for all path
indices, $l$. According to the recursive equations
in~\eqref{equation:PbPgnk}, the values of $Q(N_l,i,l)$ for all $0
\leq i \leq k $ and $1 \leq l \leq L_j$ can be calculated with the
complexity of $O(N_l k)=O\left(\frac{n}{L_j}k\right)$. According to
the recursive equations in~\eqref{equation:ENkl}, computing
$E^{\mathbf{N}}(k,l)$ requires memoization over an array of size
$O(kl)$ whose entries can be calculated with $O(k)$ operations each.
Thus, $E^{\mathbf{N}}(k,l)$ is computable with the complexity of
$O(k^2 l)$ if $Q(N_l,i,l)$'s are already given. Finally, noting that
$Q_j(n,k)=E^{\mathbf{N}}(k,L_j)$, we can compute $Q_j(n,k)$ with the
overall complexity of $O(k^2 L_j)+O\left(\frac{n}{L_j}k\right)$.
\section{Proof of Lemma V}
\label{section:ProofLemmaV} The lemma is proved by induction on $j$.
The case of $j=1$ is obviously true as
$\hat{P}_e(n,k,1)=P_e^{opt}(n,k,1)$. Let us assume this statement is
true for $j=1$ to $J-1$. Then, for $j=J$, we have
\begin{eqnarray}
&                      & \hat{P}_e(n,k,J) \nonumber \\
& \stackrel{(a)}{\leq} & \sum_{i=0}^{N_J} Q_J(N_J^{opt},i) \hat{P}_e(n-N_J^{opt},k-i,J-1) \nonumber \\
& \stackrel{(b)}{\leq} & \sum_{i=0}^{N_J} Q_J(N_J^{opt},i) P_e^{opt}(n-N_J^{opt},k-i,J-1) \nonumber \\
& \stackrel{(c)}{\leq} & \sum_{i=0}^{N_J} Q_J(N_J^{opt},i) P_e^{\mathbf{N}^{opt}}(k-i,J-1) \nonumber \\
& \stackrel{(d)}{=}    & P_e^{\mathbf{N}^{opt}}(k,J)= P_e^{opt}(n,k,J) \nonumber
\end{eqnarray}
where $\mathbf{N}^{opt}$ denotes the optimum allocation of $n$
packets among the $J$ types of paths such that the probability of
having more than $k$ lost packets is minimized. $(a)$ follows from
the recursive equation~\eqref{equation:PeNkj}, and $(b)$ is the
induction assumption. $(c)$ comes from the definition of
$P_e^{opt}(n,k,l)$, and $(d)$ is a result of
equation~\eqref{equation:PeHatnkj}.


\section{Proof of Theorem III}
\label{section:ProofTheoremIII} \textbf{Sketch of the proof:} First,
the asymptotic behavior of $Q_j(n,k)$ is analyzed, and it is shown
that for large values of $L_j$ (or equivalently $L$),
equation~\eqref{equation:Qjnk} computes the exponent of $Q_j(n,k)$
versus $L$. Next, we prove the first part of the theorem by
induction on $J$. The proof of this part is divided to two different
cases, depending on whether $\frac{K}{N}$ is larger than
$\mathbb{E}\{x_J\}$ or vice versa. Finally, the second and the third
parts of the theorem are proved by induction on $j$ while the total
number of path types, $J$, is fixed. Again, the proof is divided
into two different cases, depending on whether $\frac{K}{N}$ is
larger than $\mathbb{E}\{x_j\}$ or vice versa.


\textbf{Proof:} First, we compute the asymptotic behavior of
$Q_j(n,k)$ for $k>n\mathbb{E}\{x_j\}$, and $n$ growing
proportionally to $L_j$, i.e. $n=n'L_j$. Here, we can apply Sanov's
Theorem \cite{Dembo1998,Cover1991Sanov} as $n$ and $k$ are discrete
variables and $n'$ is a constant.

\textbf {Sanov's Theorem.} Let $X_1, X_2, \dots, X_n$ be i.i.d.
discrete random variables from an alphabet set $\mathcal X$ with the
size $| \mathcal X|$ and probability mass function (pmf) $Q(x)$. Let
$\mathcal P$ denote the set of pmf's in $\mathbb R^{| \mathcal X|}$,
i.e. $\mathcal P = \left\{ \mathbf P \in \mathbb R^{| \mathcal X|}
|~P(i) \geq 0,~\sum_{i=1}^{| \mathcal X|}{P(i)}=1 \right\}$. Also,
let $\mathcal P_L$ denote the subset of $\mathcal P$ corresponding
to all possible empirical distributions of $\mathcal X$ in $L$
observations~\cite{Cover1991Sanov}, i.e. $\mathcal P_L=\left\{ \mathbf P
\in \mathcal P |~\forall i,~LP(i) \in \mathbb Z \right\}$. For any
dense and closed set~\cite{Kelley1975} of pmf's $E \subseteq
\mathcal P$, the probability that the empirical distribution of $L$
observations belongs to the set $E$ is equal to
\begin{equation}
\mathbb P \left\{ E \right\} = \mathbb P \left\{ E \cap \mathcal P_L \right\} \doteq e^{-LD\left( \mathbf P^{\star} || \mathbf Q \right)}
\end{equation}
where $\mathbf P^{\star}= \displaystyle \argmin_{\mathbf P \in E}
D(\mathbf P || \mathbf Q)$ and $D(\mathbf P || \mathbf
Q)=\sum_{i=1}^{| \mathcal X|}{P(i) \log \frac{ P(i) }{ Q(i) } }$.

Focusing our attention on the main problem, assume that $\mathbf P$
is defined as the empirical distribution of the number of errors in
each path, i.e. for $\forall i,~1 \leq i \leq n',~P(i)$ shows the
ratio of the total paths which contain exactly $i$ lost packets.
Similarly, for $\forall i,~1 \leq i \leq n',~Q(i)$ denotes the
probability of exactly $i$ packets being lost out of the $n'$
packets transmitted on a path of type $j$. The sets $E$ and
$E_{out}$ are defined as follows
\begin{eqnarray}
E      & = & \{ \mathbf P \in \mathcal P | \sum_{i=0}^{n'} iP(i) \geq \beta \} \\ \nonumber
E_{out} & = & \{ \mathbf P \in \mathcal P | \sum_{i=0}^{n'} iP(i) = \beta \}
\end{eqnarray}
where $\beta=\dfrac{k}{n}$. Noting $E$ and $E_{out}$ are dense sets,
we can compute $Q_j(n, k)$ as
\begin{equation}
Q_j(n,k)  \stackrel{(a)}{=}  \mathbb P \left\{ E_{out} \right\}
\stackrel{(b)}{\doteq} e^{-L_j \displaystyle \min_{\mathbf P \in E_{out}} D\left( \mathbf P || \mathbf Q \right)}
\label{equation:QjnkFirst}
\end{equation}
where $(a)$ follows from the definition of $Q_j(n,k)$ as the
probability of having exactly $k$ errors out of the $n$ packets sent
over the paths of type $j$ given in
section~\ref{section:SubOptimalRateAlloc}, and $(b)$ results from
Sanov's Theorem.

Knowing the fact that the Kullback Leibler distance, $D(\mathbf P ||
\mathbf Q)$, is a convex function of $\mathbf P$ and 
$\mathbf Q$~\cite{Cover1991Distance}, we conclude that its minimum over the convex set
$E$ either lies on an interior point which is a global minimum of
the function over the whole set $\mathcal P$ or is located on the
boundary of $E$. However, we know that the global minimum of
Kullback Leibler distance occurs at $\mathbf P= \mathbf Q \notin E$.
Thus, the minimum of $D(\mathbf P || \mathbf Q)$ is located on the
boundary of $E$. This results in
\begin{eqnarray}
Q_j(n,k) & \stackrel{(a)}{\doteq} & e^{-L_j \displaystyle \min_{\mathbf P \in E_{out}} D\left( \mathbf P || \mathbf Q \right)} \nonumber \\
         & = & e^{-L_j \displaystyle \min_{\mathbf P \in E} D\left( \mathbf P || \mathbf Q \right)} \stackrel{(b)}{\doteq} e^{-\gamma_j L u_j(\dfrac{k}{n})}
\label{equation:Qjnk}
\end{eqnarray}
where $(a)$ and $(b)$ follow from
equations~\eqref{equation:QjnkFirst} and~\eqref{equation:dotEQmain},
respectively.

\vspace{\baselineskip}

\noindent 1) We prove the first part of the theorem by induction on
$J$. When $J=1$, the statement is correct for both cases of
$\frac{K}{N}>\mathbb{E}\{x_1\}$ and
$\frac{K}{N}\leq\mathbb{E}\{x_1\}$, recalling the fact that $\hat
P_e(n, k, 1) = P_e^{opt}(n, k, 1)$ and $u_1(x)=0$ for $x \leq
\mathbb{E}\{x_1\}$. Now, let us assume the first part of the theorem
is true for $j=1$ to $J-1$. We prove the same statement for $J$ as
well. The proof can be divided into two different cases, depending
on whether $\frac{K}{N}$ is larger than $\mathbb{E}\{x_J\}$ or vice
versa.

\subsection*{1.1) $\dfrac{K}{N}>\mathbb{E}\{x_J\}$}
According to the definition, the value of $\hat P_e(N, K, J)$ is
computed by minimizing $\sum_{i=0}^{n_J}Q_J(n_J,i)$$\hat{P}_e(N-n_J,K-i,J-1)$ over $n_J$
(see equation~\eqref{equation:PeHatnkj}). Now, we show that for any
value of $n_J$, the corresponding term in the minimization is
asymptotically at least equal to $P_e^{opt}(N, K, J)$. $n_J$ can
take integer values in the range $0 \leq n_J \leq N$. We split this
range into three non-overlapping intervals of $0\leq n_J \leq
\epsilon L$, $\epsilon L \leq n_J \leq N ( 1 - \epsilon)$, and
$N(1-\epsilon) < n_J \leq N$ for any arbitrary constant $\epsilon
\leq \min \left\{ \gamma_j, 1 - \frac{K}{N} \right\}$. The reason is
that equation~\eqref{equation:Qjnk} is valid in the second interval
only, and we need separate analyses for the first and last
intervals.

First, we show the statement for $\epsilon L \leq n_J \leq N ( 1 -
\epsilon)$. Defining $i_J = \lfloor n_J \frac{K}{N} \rfloor$, we
have
\begin{eqnarray}
\frac{i_J}{n_J} = \frac{K}{N} + O(\frac{1}{L}), \nonumber \\
\frac{K-i_J}{N-n_J} = \frac{K}{N} + O(\frac{1}{L})
\end{eqnarray}
as $\epsilon$ is constant, and $K=O(L)$, $N=O(L)$. Hence, we have
\begin{eqnarray}
& & \sum_{i=0}^{n_J}{Q_J(n_J,i)\hat{P}_e(N-n_J,K-i,J-1)} \nonumber \\
& \geq & Q_J(n_J,i_J)\hat{P}_e(N-n_J,K-i_J,J-1) \nonumber \\
& \stackrel{(a)}{\doteq} & \displaystyle e^{-L \displaystyle\sum_{j=1}^{J}\gamma_j u_j \left( \frac{K}{N} + O \left( \frac{1}{L} \right) \right) } \nonumber \\
& \stackrel{(b)}{\doteq} & \displaystyle e^{-L \displaystyle\sum_{j=1}^{J}\gamma_j u_j \left( \frac{K}{N} \right) }
\label{equation:SumHatPeNKJequal}
\end{eqnarray}
where $(a)$ follows from~\eqref{equation:Qjnk} and the induction
assumption, and $(b)$ follows from the fact that $u_j()$'s are
differentiable functions according to Lemma I in
subsection~\ref{subsection:IdenticalPaths}.

For $0\leq n_J \leq \epsilon L$, since $\epsilon < \gamma_j$, the
number of packets assigned to the paths of type $J$ is less than the
number of such paths. Thus, one packet is allocated to $n_J$ of the
paths, and the rest of the paths of type $J$ are not used. Defining
$\pi_{b,J}$ as the probability of a path of type $J$ being in the
bad state, we can write
\begin{equation}
Q_J(n_J,n_J)=\pi_{b,J}^{n_J}=\displaystyle e^{-n_J \log \left( \dfrac{1}{\pi_{b,J}} \right)}.
\end{equation}
Therefore, for $0\leq n_J \leq \epsilon L$, we have
\begin{eqnarray}
& & \sum_{i=0}^{n_J}{Q_J(n_J,i)\hat{P}_e(N-n_J,K-i,J-1)} \nonumber \\
& \geq & Q_J(n_J,n_J)\hat{P}_e(N-n_J,K-n_J,J-1) \nonumber \\
& \doteq & \displaystyle e^{ \displaystyle -L \displaystyle\sum_{j=1}^{J-1} \gamma_j u_j \left( \dfrac{K-n_J}{N-n_J} \right) -n_J \log \left( \dfrac{1}{\pi_{b,J}} \right) } \nonumber \\
& \stackrel{(a)}{\geq} & \displaystyle e^{ -L \displaystyle\sum_{j=1}^{J-1} \gamma_j u_j \left( \frac{K}{N} \right)  - \displaystyle L \epsilon \log \left( \dfrac{1}{\pi_{b,J}} \right) } \nonumber \\
& \stackrel{(b)}{\doteq} & e^{ \displaystyle -L \displaystyle\sum_{j=1}^{J-1} \gamma_j u_j \left( \frac{K}{N} \right) } \geq  e^{ \displaystyle-L \displaystyle\sum_{j=1}^{J} \gamma_j u_j \left( \frac{K}{N} \right) }
\label{equation:SumHatPeNKJless}
\end{eqnarray}
where $(a)$ follows from the fact that $\frac{K-n_J}{N-n_J} \leq
\frac{K}{N}$, and $(b)$ results from the fact that we can select
$\epsilon$ arbitrarily small.

Finally, we prove the statement for the case $n_J > N(1-\epsilon)$.
In this case, we have
\begin{eqnarray}
& & \sum_{i=0}^{n_J}{Q_J(n_J,i)\hat{P}_e(N-n_J,K-i,J-1)} \nonumber \\
& \geq & Q_J(n_J, K)\hat{P}_e(N-n_J,0,J-1) \nonumber \\
& \stackrel{(a)}{\geq} & \displaystyle e^{ \displaystyle -L \gamma_J u_J \left( \dfrac{K}{N \left( 1- \epsilon \right)} \right)} \nonumber \\
& \stackrel{(b)}{\dot \geq} & e^{ \displaystyle-L \displaystyle\sum_{j=1}^{J} \gamma_j u_j \left( \frac{K}{N} \right) }
\label{equation:SumHatPeNKJmore}
\end{eqnarray}
where $(a)$ follows from the fact that $\epsilon < 1 - \frac{K}{N}$
and $\hat P_e(n, 0, j)=1,$ for all $n$ and $j$. Setting $\epsilon$
small enough results in $(b)$.

Inequalities~\eqref{equation:SumHatPeNKJequal},~\eqref{equation:SumHatPeNKJless},
and~\eqref{equation:SumHatPeNKJmore} result in
\begin{eqnarray}
\hat{P}_e(N,K,J) & \dot{\geq} & e^{ -L \displaystyle\sum_{j=1}^{J} \gamma_j u_j \left( \alpha \right) }
\label{equation:HatPeNKJmorePeOptNKJ}
\end{eqnarray}
Combining~\eqref{equation:HatPeNKJmorePeOptNKJ} with Lemma V proves
the first part of Theorem III for the case when
$\frac{K}{N}>\mathbb{E}\{x_J\}$.

\subsection*{1.2) $\dfrac{K}{N}\leq\mathbb{E}\{x_J\}$}
Similar to the case of $\dfrac{K}{N}>\mathbb{E}\{x_J\}$ in
subsection 1.1, we show that for any value of $0 \leq n_J \leq N$,
the corresponding term of the minimization in
equation~\eqref{equation:PeHatnkj} is asymptotically at least equal
to $P_e^{opt}(N, K, J)$. Again, the range of $n_J$ is partitioned
into three non-overlapping intervals.

For any arbitrary $0<\epsilon< \min \left\{ \gamma_J, 1 -
\frac{K}{N} , \frac{1}{K} \right\}$, and for all $n_J$ in the range
of $\epsilon L < n_J \leq N ( 1 - \epsilon ) $, we define $i_J$ as
$i_J = \lceil n_J \mathbb{E}\{x_J\} \rceil$. We have
\begin{eqnarray}
\frac{i_J}{n_J} & = & \mathbb{E}\{x_J\} + O \left( \frac{1}{L} \right) \geq \mathbb{E}\{x_J\} \nonumber \\
\frac{K-i_J}{N-n_J} & < & \frac{K}{N} + O \left( \frac{1}{L} \right)
\label{equation:K_iJoverN_nJ}
\end{eqnarray}
Hence,
\begin{eqnarray}
& & \sum_{i=0}^{n_J}{Q_J(n_J,i)\hat{P}_e(N-n_J,K-i,J-1)} \nonumber \\
& \geq & Q_J(n_J,i_J)\hat{P}_e(N-n_J,K-i_J,J-1) \nonumber \\
& \stackrel{(a)}{\doteq} & e^{\displaystyle -L\gamma_J u_J \left( \dfrac{i_J}{n_J}\right) -L \displaystyle\sum_{j=1}^{J-1} \gamma_j u_j \left(\dfrac{K-i_J}{N-n_J}\right) } \nonumber \\
& \stackrel{(b)}{\geq} & e^{\displaystyle -L\gamma_J u_J \left( \mathbb{E}\{x_J\}+ O\left(\frac{1}{L}\right) \right) } \cdot \nonumber \\
& &   e^{\displaystyle -L \displaystyle\sum_{j=1}^{J-1} \gamma_j u_j \left( \frac{K}{N} + O \left( \frac{1}{L} \right) \right) } \nonumber \\
& \stackrel{(c)}{\doteq} & \displaystyle e^{-L \displaystyle\sum_{j=1}^{J} \gamma_j u_j \left( \frac{K}{N} \right) }
\label{equation:SumHatPeNKJequal2}
\end{eqnarray}
where $(a)$ follows from~\eqref{equation:Qjnk} and the induction
assumption, and $(b)$ is based on~\eqref{equation:K_iJoverN_nJ}.
$(c)$ results from the facts that $u_j()$'s are differentiable
functions, and we have $u_J \left( \mathbb{E}\{x_J\} \right)=0$,
both according to Lemma I in
subsection~\ref{subsection:IdenticalPaths}.

For $0\leq n_J \leq \epsilon L$, the analysis of section 1.1 and
inequality~\eqref{equation:SumHatPeNKJless} are still valid. For
$n_J > ( 1 - \epsilon ) N$, we set $i_J = \lceil \mathbb E \left\{
x_J \right\} n_J \rceil$. Now, we have
\begin{equation}
i_J \geq n_J \mathbb{E}\{x_J\} > (1-\epsilon) N \mathbb{E}\{x_J\} \geq (1-\epsilon) K.
\end{equation}
The above inequality can be written as
\begin{equation}
K-i_J < \epsilon K < 1
\label{equation:KminusiJ}
\end{equation}
since $\epsilon< \frac{1}{K}$. Noting that $K$ and $i_J$ are integer
values, it is concluded that $K \leq i_J$. Now, we can write
\begin{eqnarray}
& & \sum_{i=0}^{n_J}{Q_J(n_J,i)\hat{P}_e(N-n_J,K-i,J-1)} \nonumber \\
& \geq & Q_J(n_J, i_J)\hat{P}_e(N-n_J, K - i_J,J-1) \nonumber \\
& \stackrel{(a)}{=} & Q_J(n_J, i_J) \nonumber \\
& \dot\geq & \displaystyle e^{ \displaystyle -L \gamma_J u_J \left( \mathbb E \left\{ x_J \right\} + \frac{1}{n_J} \right)} \nonumber \\
& \stackrel{(b)}{\dot\geq} & \displaystyle e^{ \displaystyle -L \gamma_J u_J \left( \mathbb E \left\{ x_J \right\} + \frac{1}{\left( 1 - \epsilon \right) N} \right)} \nonumber \\
& \doteq &  e^{ \displaystyle -L \gamma_J u_J \left( \mathbb E \left\{ x_J \right\} + O\left(\frac{1}{L}\right) \right)} \stackrel{(c)}{\doteq} 1
\label{equation:SumHatPeNKJmore2}
\end{eqnarray}
where $(a)$ follows from the fact that $K \leq i_J$, and $\hat
P_e(n, k, j)=1,$ for $k \leq 0$. $(b)$ and $(c)$ result from $n_J >
(1 - \epsilon) N$ and $u_J \left( \mathbb{E}\{x_J\} \right)=0$,
respectively.

Hence, inequalities~\eqref{equation:SumHatPeNKJless}, \eqref{equation:SumHatPeNKJequal2}, and \eqref{equation:SumHatPeNKJmore2} result in
\begin{eqnarray}
\hat{P}_e(N,K,J) & \dot{\geq} & e^{-L \displaystyle\sum_{j=1}^{J}
\gamma_j u_j \left( \alpha \right) }
\end{eqnarray}
which proves the first part of Theorem III for the case of
$\frac{K}{N}\leq \mathbb{E}\{x_J\}$ when combined with Lemma V.

\vspace{\baselineskip}

\noindent 2) We prove the second and the third parts of the theorem
by induction on $j$ while the total number of types, $J$, is fixed.
The proof of the statements for the base of the induction, $j=J$, is
similar to the proof of the induction step, from $j+1$ to $j$.
Hence, we just give the proof for the induction step. Assume the
second and the third parts of the theorem are true for $m=J$ to
$j+1$. We prove the same statements for $j$. The proof is divided
into two different cases, depending on whether $\frac{K}{N}$ is
larger than $\mathbb{E}\{x_j\}$ or vice versa.

Before we proceed further, it is helpful to introduce two new
parameters $N'$ and $K'$ as
\begin{eqnarray}
N' & = & N-\sum_{m=j+1}^{J}\hat{N}_j \nonumber \\
K' & = & K-\sum_{m=j+1}^{J}K_j . \nonumber
\end{eqnarray}
According to the above definitions and the induction assumptions, it is obvious that
\begin{equation}
\dfrac{K'}{N'} = \dfrac{K}{N} + o(1)= \alpha + o(1).
\label{equation:KprimeNprime}
\end{equation}

\subsection*{2.1) $\dfrac{K}{N}>\mathbb{E}\{x_j\}$}
First, by contradiction, it will be shown that for small enough
values of $\epsilon >0$, we have $\hat{N}_j>\epsilon N'$. Let us
assume the opposite is true, i.e. $\hat{N}_j\leq\epsilon N'$. Then,
we can write
\begin{eqnarray}
& & \hat{P}_e(N',K',j) \nonumber \\
& \stackrel{(a)}{=} &  \sum_{i=0}^{\hat{N}_j}\hat{P}_e(N'-\hat{N}_j,K'-i,j-1)Q_j(\hat{N}_j,i)   \nonumber \\
&       \geq        & \hat{P}_e(N'-\hat{N}_j,K'-\hat{N}_j,j-1)Q_j(\hat{N}_j,\hat{N}_j)  \nonumber \\
& \stackrel{(b)}{\doteq} & Q_j(\hat{N}_j,\hat{N}_j) e^{ \displaystyle-L \displaystyle\sum_{r=1}^{j-1} \gamma_r u_r \left( \dfrac{K' - \hat{N}_j}{N'-\hat{N}_j} \right) } \nonumber \\
& \stackrel{(c)}{\geq} & e^{ \displaystyle - L n_0 \left( 1 - \sum_{r=j+1}^{J} \eta_r \right) \epsilon \log \left( \dfrac{1}{\pi_{b,j}} \right)} \cdot \nonumber \\
& & e^{ -L \displaystyle\sum_{r=1}^{j-1} \gamma_r u_r \left( \dfrac{K'}{N'} \right) } \nonumber \\
& \stackrel{(d)}{\dot >} &  e^{ \displaystyle -L \displaystyle\sum_{r=1}^{j} \gamma_r u_r \left( \alpha \right) }
\label{equation:HatPeNKjlessExponent}
\end{eqnarray}
where $(a)$ follows from equation~\eqref{equation:PeHatnkj} and step
(2) of our suboptimal algorithm, $(b)$ results from the first part
of Theorem III, and $(c)$ can be justified using arguments similar
to those of inequality~\eqref{equation:SumHatPeNKJless}. $(d)$ is
obtained assuming $\epsilon$ is small enough such that the
corresponding term in the exponent is strictly less than $L \gamma_j
u_j \left( \frac{K'}{N'} \right)$ and also the fact that
$\frac{K'}{N'} = \alpha + o(1)$. The result
in~\eqref{equation:HatPeNKjlessExponent} is obviously in
contradiction with the first part of Theorem III, proving that
$\hat{N}_j>\epsilon N'$.

Now, we show that if $\hat N_j > (1 - \epsilon) N'$ for arbitrarily
small values of $\epsilon$, we should have $\mathbb E \left\{ x_r
\right\} > \alpha$ for all $1 \leq r \leq j-1$. In such a case, we
observe $\frac{\hat N_j}{N'}=1+o(1)$, proving the second statement
of Theorem III. To show this, let us assume $\hat N_j > (1 -
\epsilon) N'$. Hence,
\begin{eqnarray}
\hat{P}_e(N',K',j) & \hspace{-0.3cm} = &  \hspace{-0.3cm} \sum_{i=0}^{\hat{N}_j}\hat{P}_e(N'-\hat{N}_j,K'-i,j-1)Q_j(\hat{N}_j,i)   \nonumber \\
&     \hspace{-0.3cm} \dot \geq        & \hspace{-0.3cm} \hat{P}_e(N'-\hat{N}_j,0,j-1)Q_j(\hat{N}_j,K')  \nonumber \\
& \hspace{-0.3cm} \stackrel{(a)}{\dot\geq} & \hspace{-0.3cm} e^{-L \gamma_j u_j \left( \frac{K'}{ \left( 1 - \epsilon \right) N'} \right) }
\stackrel{(b)}{\doteq}  e^{-L \gamma_j u_j \left( \alpha + o(1) \right)}
\label{equation:HatPeNKjmoreExponent}
\end{eqnarray}
where $(a)$ follows from the fact that $\hat P_e(n,0,j)=1$, for all
values of $n$ and $j$, and the fact that $\hat N_j \geq (1 -
\epsilon) N'$. $(b)$ is obtained by making $\epsilon$ arbitrarily
small and using equation~\eqref{equation:KprimeNprime}. Applying
(\ref{equation:HatPeNKjmoreExponent}) and knowing the fact that
$\hat P_e(N',K',j) \doteq e^{-L\sum_{r=1}^{j} \gamma_r u_r \left(
\alpha \right) }$, we conclude that $\mathbb E \left\{ x_r \right\}
> \alpha$, for all values of $1 \leq r \leq j-1$.

$\hat{P}_e(N',K',j)$ can be written as
\begin{eqnarray}
& & \hat{P}_e(N',K',j) \nonumber \\
& = &  \min_{ 0 \leq N_j \leq N'} \sum_{i=0}^{N_j}\hat{P}_e(N'-N_j,K'-i,j-1)Q_j(N_j,i) \nonumber \\
& \stackrel{(a)}{\doteq} & \min_{ \epsilon N' \leq N_j \leq (1-\epsilon) N'} ~~\max_{0 \leq i \leq N_j} \nonumber \\
& & \hat{P}_e(N'-N_j,K'-i,j-1)Q_j(N_j,i) \nonumber \\
& \stackrel{(b)}{\doteq} & \min_{ \epsilon N' \leq N_j \leq (1-\epsilon) N'} ~~\max_{\mathbb E \{ x_j \}N_j < i \leq N_j} \nonumber \\
& & e^{ \displaystyle-L \gamma_j u_j \left( \dfrac{i}{N_j} \right) -L \displaystyle\sum_{r=1}^{j-1} \gamma_r u_r \left( \dfrac{K'-i}{N'-N_j} \right) } \nonumber \\
& \doteq & e^{ \displaystyle-L \max_{\epsilon N' \leq N_j \leq (1-\epsilon) N'} ~~\min_{\mathbb E \{ x_j \}N_j < i \leq N_j} M_d(i,N_j)} \nonumber \\
& \stackrel{(c)}{\doteq} & e^{ \displaystyle-L \max_{\epsilon \leq \lambda_j \leq (1-\epsilon) }~~ \min_{\mathbb E \{ x_j \} \lambda_j < \beta_j \leq \lambda_j} M_c(\beta_j,\lambda_j)}.
\label{equation:HatPeNKjExponent}
\end{eqnarray}
where $M_d(i,N_j)$ and $M_c(\beta_j, \lambda_j)$ are defined as
\begin{eqnarray}
M_d(i,N_j) & = & \gamma_j u_j \left( \dfrac{i}{N_j} \right) + \displaystyle\sum_{r=1}^{j-1} \gamma_r u_r \left( \dfrac{K'-i}{N'-N_j} \right) \nonumber \\
M_c(\beta_j,\lambda_j) & = & \gamma_j u_j \left( \dfrac{\beta_j}{\lambda_j} \right) + \displaystyle\sum_{r=1}^{j-1} \gamma_r u_r \left( \dfrac{\alpha - \beta_j}{1-\lambda_j} \right). \nonumber
\end{eqnarray}
In~\eqref{equation:HatPeNKjExponent}, $(a)$ follows from the fact
that $\hat N_j$ is bounded as $\epsilon N' \leq \hat N_j \leq (1 -
\epsilon) N'$. $(b)$ results from equation~\eqref{equation:Qjnk},
$\hat{P}_e(n,k,j)$ being a decreasing function of $k$, and the fact
that we have $Q_j(N_j, i) \leq 1 \doteq Q_j(N_j, \mathbb E \left\{
x_j \right\} N_j)$  for $i<\mathbb E \left\{ x_j \right\} N_j$.
$\beta_j$ and $\lambda_j$ are defined as $\beta_j= \frac{i}{N'}$ and
$\lambda_j=\frac{N_j}{N'}$. $(c)$ is a result of having
$M_c(\beta_j, \lambda_j)=M_d(i,N_j)+O\left(\frac{1}{L}\right)$.
Hence, the discrete to continuous relaxation is valid.

Let us define $\left(\beta_j^* , \lambda_j^* \right)$ as the values
of $\left( \beta_j , \lambda_j \right)$ which solve the max-min
problem in \eqref{equation:HatPeNKjExponent}. Differentiating
$M_c(\beta_j, \lambda_j)$ with respect to $\beta_j$ and $\lambda_j$
results in
\begin{eqnarray}
0&\hspace{-0.3cm}=&\hspace{-0.3cm}\dfrac{\gamma_j}{\lambda_j^*} l_j \left( \dfrac{\beta_j^*}{\lambda_j^*} \right)
- \sum_{ \substack{ r=1,\\ \mathbb E \{x_r\} < \zeta } }^{j-1} \dfrac{\gamma_r}{1-\lambda_j^*} l_r \left( \zeta \right) \nonumber \\
0&\hspace{-0.3cm}=&\hspace{-0.3cm}\left\{ -\dfrac{\gamma_j \beta_j^*}{\lambda_j^{*2}} l_j \left( \dfrac{\beta_j^*}{\lambda_j^*} \right) + \sum_{ \substack{ r=1, \\ \mathbb E \{x_r\} < \zeta} }^{j-1} \dfrac{\gamma_r (\alpha-\beta_j^*)}{(1-\lambda_j^*)^2} l_r \left( \zeta \right) \right. \nonumber \\
&\hspace{-0.3cm}&\hspace{-0.3cm}\left. + \left( \dfrac{\gamma_j}{\lambda_j^*} l_j \left( \dfrac{\beta_j^*}{\lambda_j^*} \right) - \sum_{ \substack{ r=1,\\ \mathbb E \{x_r\} < \zeta } }^{j-1} \dfrac{\gamma_r}{1-\lambda_j^*} l_r \left( \zeta \right) \right) \dfrac{\partial \beta_j^*}{\partial \lambda_j} |_{\lambda_j=\lambda_j^*} \right\}  \nonumber
\end{eqnarray}
where $\zeta=\dfrac{\alpha-\beta_j^*}{1-\lambda_j^*}$. Solving the
above equations gives the unique optimum solution $(\beta_j^*,
\lambda_j^*)$ as
\begin{eqnarray}
\beta_j^* & = & \alpha \lambda_j^* \nonumber \\
\lambda_j^* & = & \dfrac {\gamma_j l_j(\alpha)} {\displaystyle\sum_{r=1,\alpha>\mathbb{E}\{x_r\}}^{j} l_r(\alpha)}
\end{eqnarray}
Hence, the integer parameters $K_j, \hat N_j$ defined in the
suboptimal algorithm have to satisfy $\frac{K_j}{N'}=\beta_j^*+o(1)$
and $\frac{\hat N_j}{N'}=\lambda_j^*+o(1)$, respectively. Based on
the induction assumption, it is easy to show that
\begin{equation}
\frac{N'}{N}=\frac { \displaystyle \sum_{r=1, \mathbb E \left\{ x_r
\right\} < \alpha}^{j} \gamma_r u_r(\alpha) }{ \displaystyle
\sum_{r=1, \mathbb E \left\{ x_r \right\} < \alpha}^{J} \gamma_r
u_r(\alpha)} \label{equation:NprimeN}
\end{equation}
which completes the proof for the case of
$\mathbb E \left\{ x_j \right\} < \frac{K}{N} $.

\subsection*{2.2) $\dfrac{K}{N}\leq\mathbb{E}\{x_j\}$}
In this case, we show that $\frac{\hat N_j}{N}=o(1)$. Defining
$i_j=\lceil \mathbb{E}\{x_j\} \hat N_j \rceil$, we have
\begin{equation}
\dfrac{K'-i_j}{N'-\hat{N}_j}=\alpha - (\mathbb{E}\{x_j\} - \alpha) \frac{\hat N_j}{N' - \hat N_j} + o(1)
\label{equation:K_ijoverN_hatNj}
\end{equation}
using equation~\eqref{equation:KprimeNprime}. Now, we have
\begin{eqnarray}
& & \hat{P}_e(N',K',j) \nonumber \\
& = &  \sum_{i=0}^{\hat{N}_j}\hat{P}_e(N'-\hat{N}_j,K'-i,j-1)Q_j(\hat{N}_j,i) \nonumber \\
& \geq & \hat{P}_e(N'-\hat{N}_j,K'-i_j,j-1)Q_j(\hat{N}_j,i_j) \nonumber \\
& \stackrel{(a)}{\doteq} & e^{\displaystyle -L\gamma_j u_j \left( \mathbb{E}\{x_j\} + o(1) \right) } \cdot \nonumber \\
& & e^{-L \displaystyle\sum_{r=1}^{j-1} \gamma_r u_r \left( \alpha - (\mathbb{E}\{x_j\} - \alpha) \frac{\hat N_j}{N' - \hat N_j} \right) } \nonumber \\
& \doteq & e^{\displaystyle -L \displaystyle\sum_{r=1}^{j-1} \gamma_r u_r \left( \alpha - (\mathbb{E}\{x_j\} - \alpha) \frac{\hat N_j}{N' - \hat N_j} \right) }
\label{equation:HatPeNKjExponent2}
\end{eqnarray}
where $(a)$ follows from the first part of Theorem III
and~\eqref{equation:Qjnk}. On the other hand, according to the
result of the first part of Theorem III, we know that
\begin{equation}
\hat{P}_e(N',K',j)  \doteq e^{\displaystyle -L \displaystyle\sum_{r=1}^{j-1} \gamma_r u_r \left( \alpha \right) }. \label{equation:HatPeNKjExponent3}
\end{equation}
According to Lemma I, $u_r(\beta)$ is an increasing function of
$\beta$ for all $1 \leq r \leq j-1$. Thus, $\sum_{r=1}^{j-1}
\gamma_r u_r \left( \beta \right) $ is also a one-to-one increasing
function of $\beta$. Noting this fact and comparing
(\ref{equation:HatPeNKjExponent2}) and
(\ref{equation:HatPeNKjExponent3}), we conclude that $\frac{\hat
N_j}{N'} = o(1)$ as $\mathbb{E} \left\{ x_j \right\} - \alpha$ is
strictly positive. Noting (\ref{equation:NprimeN}), we have
$\frac{\hat N_j}{N} = o(1)$ which proves the second part of Theorem
III for the case of $\frac{K}{N}\leq\mathbb{E}\{x_j\}$.

\bibliographystyle{IEEEtran}
\bibliography{TechRep4}

\begin{thebibliography}{10}
\providecommand{\url}[1]{#1}
\csname url@samestyle\endcsname
\providecommand{\newblock}{\relax}
\providecommand{\bibinfo}[2]{#2}
\providecommand{\BIBentrySTDinterwordspacing}{\spaceskip=0pt\relax}
\providecommand{\BIBentryALTinterwordstretchfactor}{4}
\providecommand{\BIBentryALTinterwordspacing}{\spaceskip=\fontdimen2\font plus
\BIBentryALTinterwordstretchfactor\fontdimen3\font minus
  \fontdimen4\font\relax}
\providecommand{\BIBforeignlanguage}[2]{{%
\expandafter\ifx\csname l@#1\endcsname\relax
\typeout{** WARNING: IEEEtran.bst: No hyphenation pattern has been}%
\typeout{** loaded for the language `#1'. Using the pattern for}%
\typeout{** the default language instead.}%
\else
\language=\csname l@#1\endcsname
\fi
#2}}
\providecommand{\BIBdecl}{\relax}
\BIBdecl

\bibitem{Bolot1999}
{ J.C. Bolot, S. Fosse-Parisis, and D. Towsley}, ``{Adaptive FEC-based error
  control for Internet telephony},'' in \emph{IEEE INFOCOM, Proc. IEEE Vol. 3},
  1999, pp. 1453--1460.

\bibitem{Bolot1996}
{J.C. Bolot and T. Turletti}, ``{Adaptive Error Control For Packet Video In The
  Internet},'' in \emph{Proc. IEEE International Conference on Image
  Processing}, 1996, pp. 25 -- 28.

\bibitem{Nguyen2003}
{ T. Nguyen and A. Zakhor }, ``{Path diversity with forward error correction
  (pdf) system for packet switched networks},'' in \emph{IEEE INFOCOM Proc.
  IEEE Vol. 1}, 2003, pp. 663-- 672.

\bibitem{Nguyen2004}
{T. Nguyen and A. Zakhor}, ``{Multiple Sender Distributed Video Streaming},''
  \emph{IEEE transactions on multimedia}, vol.~6, no.~2, pp. 315-- 326, 2004.

\bibitem{Leannec1999}
{F. L. Leannec, F. Toutain, and C. Guillemot}, ``{Packet Loss Resilient MPEG-4
  Compliant Video Coding for the Internet},'' \emph{Journal of Image
  Communication, Special Issue on Real-time video over the Internet}, no.~15,
  pp. 35--56, 1999.

\bibitem{Fashandi2008isit}
{S. Fashandi, S. Oveisgharan, and A.K. Khandani}, ``{Coding over an Erasure
  Channel with a Large Alphabet Size},'' in \emph{IEEE International Symposium
  on Information Theory, ISIT '08}, 2008.

\bibitem{Fashandi20083}
------, ``{Coding over an Erasure Channel with a Large Alphabet Size},'' 2008,
  library and Archives Canada Technical Report UW-ECE \#2008-06,
  \url{http://cst.uwaterloo.ca/r/2008-06\_Shervan.pdf}.

\bibitem{Han20062}
{H. Han, S. Shakkottai, C.V. Hollot, R. Srikant, and D. Towsley}, ``{Multi-Path
  TCP: A Joint Congestion Control and Routing Scheme to Exploit Path Diversity
  in the Internet},'' \emph{IEEE/ACM Transactions on Networking}, vol.~14,
  no.~6, pp. 1260 -- 1271, 2006.

\bibitem{Mao2005}
{ S. Mao, S.S. Panwar, and Y.T. Hou}, ``{On optimal partitioning of realtime
  traffic over multiple paths},'' in \emph{INFOCOM 2005, Proc. IEEE Vol. 4},
  2005, pp. 2325--2336.

\bibitem{Fashandi2007Glob}
{S. Fashandi, S. Oveisgharan, and A.K. Khandani}, ``{Path Diversity in Packet
  Switched Networks: Performance Analysis and Rate Allocation},'' in \emph{IEEE
  Global Telecommunications Conference, GLOBECOM '07}, 2007, pp. 1840--1844.

\bibitem{Han2006}
{J. Han, D. Watson, and F. Jahanian}, ``{An Experimental Study of Internet Path
  Diversity},'' \emph{IEEE Transactions on Dependable and Secure Computing},
  vol.~3, no.~4, pp. 273 -- 288, 2006.

\bibitem{Han2004}
{J. Han and F. Jahanian}, ``{Impact of Path Diversity on Multi-homed and
  Overlay Networks},'' in \emph{International Conference on Dependable Systems
  and Networks}, 2004, pp. 29--38.

\bibitem{Spring2004}
{N. Spring, R. Mahajan, D. Wetherall, and T. Anderson}, ``{Measuring ISP
  Topologies with Rocketfuel},'' \emph{IEEE/ACM Transactions on Networking},
  vol.~12, no.~1, pp. 2-- 16, 2004.

\bibitem{Teixeira2003}
{R. Teixeira, K. Marzullo, S. Savage, and G. M. Voelker}, ``{In Search of Path
  Diversity in ISP Networks},'' in \emph{Proceedings of the 3rd ACM SIGCOMM
  Conference on Internet Measurement}, 2003, pp. 313 -- 318.

\bibitem{Barbasi1999}
{A. L. Barbasi and R. Albert}, ``{Emergence of Scaling in Random Networks},''
  \emph{Science}, vol. 286, no. 5439, pp. 509--512, 1999.

\bibitem{Andersen2001}
{David G. Andersen}, \emph{{Resilient Overlay Networks}}.\hskip 1em plus 0.5em
  minus 0.4em\relax Master's Thesis, Massachusetts Institute of Technology,
  2001.

\bibitem{Liang2001}
{ Y. J. Liang, E. G. Steinbach, and B. Girod }, ``{Multi-stream Voice over IP
  using Packet Path Diversity},'' in \emph{IEEE Fourth Workshop on Multimedia
  Signal Processing}, 2001, pp. 555--560.

\bibitem{Nelakuditi2004}
{S. Nelakuditi, Z. Zhang, and D. H. C. Du}, ``{On Selection of Candidate Paths
  for Proportional Routing},'' \emph{Elsevier Computer Networks}, vol.~44,
  no.~1, pp. 79--102, 2004.

\bibitem{Andersen2003}
{D. G. Andersen, A. C. Snoeren, and H. Balakrishnan}, ``{Best-path vs.
  Multi-path Overlay Routing},'' in \emph{Proceedings of the 3rd ACM SIGCOMM
  Conference on Internet Measurement}, 2003, pp. 91 -- 100.

\bibitem{Chun2004}
{B-G Chun, R. Fonseca, I. Stoica, and J. Kubiatowicz}, ``{Characterizing
  Selfishly Constructed Overlay Routing Networks},'' in \emph{IEEE INFOCOM},
  2004, pp. 1329--1339.

\bibitem{Han2005}
{J. Han, D. Watson, and F. Jahanian}, ``{Topology Aware Overlay Networks},'' in
  \emph{IEEE INFOCOM}, vol.~4, 2005, pp. 2554-- 2565.

\bibitem{Guo2003}
{M. Guo, Q. Zhang, and W. Zhu }, ``{Selecting Path-diversified Servers in
  Content Distribution Networks},'' in \emph{IEEE Global Telecommunications
  Conference, GLOBECOM '03}, vol.~6, 2003, pp. 3181--3185.

\bibitem{Akella2008}
{A. Akella, B. Maggs, S. Seshan, and A. Shaikh}, ``{On the Performance Benefits
  of Multihoming Route Control},'' \emph{IEEE/ACM Transactions on Networking},
  vol.~16, no.~1, pp. 91--104, 2008.

\bibitem{Akella2004}
{A. Akella, J. Pang, B. Maggs, S. Seshan, and A. Shaikh}, ``{A Comparison of
  Overlay Routing and Multihoming Route Control},'' in \emph{ACM SIGCOMM},
  2004, pp. 93 -- 106.

\bibitem{Srinivasan2007}
{S. Srinivasan}, \emph{{Design and Use of Managed Overlay Networks}}.\hskip 1em
  plus 0.5em minus 0.4em\relax PhD Dissertation, Georgia Institute of
  Technology, 2007.

\bibitem{Cha2006}
{M. Cha, S. Moon, C. D. Park, and A. Shaikh}, ``{Placing Relay Nodes for
  Intra-Domain Path Diversity},'' in \emph{IEEE INFOCOM}, 2006, pp. 1--12.

\bibitem{Eppstein1994}
{ David Eppstein}, ``{Finding the $k$ shortest paths},'' in \emph{Proc. 35th
  Symp. Foundations of Computer Science}, 1994, pp. 154--165.

\bibitem{Ogier1993}
{Richard G. Ogier, Vlad Rutenburg, and Nauchum Shacham}, ``{Distributed
  Algorithms for Computing Shortest Pairs of Disjoint Paths},'' \emph{IEEE
  transactions on information theory}, vol.~39, no.~2, pp. 443-- 455, 1993.

\bibitem{Clarck2006}
{D. Clark, W. Lehr, S. Bauer, P. Faratin, R. Sami, and J. Wroclawski},
  ``{Overlay Networks and Future of the Internet},'' \emph{Journal of
  Communications and Strategies}, vol.~3, no.~63, pp. 1--21, 2006.

\bibitem{Cha2007}
{M. Cha}, \emph{{Network Support for Emerging Multimedia Streaming
  Services}}.\hskip 1em plus 0.5em minus 0.4em\relax PhD Dissertation, Korea
  Advanced Institute of Science and Technology, 2007.

\bibitem{Karrer2003}
{ Roger Karrer, and Thomas Gross}, ``{Multipath Streaming in Best-Effort
  Networks},'' in \emph{Proc. of the IEEE International Conference on
  Communications (ICC'03)}, 2003.

\bibitem{Apostolopoulos2002}
{ J.G. Apostolopoulos, T. Wong, W. Tan, and S.J. Wee}, ``{On Multiple
  Description Streaming with Content Delivery Networks},'' in \emph{IEEE
  INFOCOM, Proc. IEEE Vol. 3}, 2002, pp. 1736 -- 1745.

\bibitem{AkamaiSureRoute}
``{Akamai SureRoute},''
  \url{http://www.akamai.com/dl/feature\_sheets/fs\_edge\\suite\_sureroute.pdf%
}.

\bibitem{Ghanassi2006}
{M. Ghanassi and P. Kabal}, ``{Optimizing Voice-over-IP Speech Quality Using
  Path Diversity},'' in \emph{IEEE 8th Workshop on Multimedia Signal
  Processing}, 2006, pp. 155--160.

\bibitem{Chakareski2003}
{ J. Chakareski and B. Girod}, ``{Rate-distortion optimized packet scheduling
  and routing for media streaming with path diversity},'' in \emph{Proc. IEEE
  Data Compression Conference}, 2003, pp. 203-- 212.

\bibitem{Afergan2006}
{M. Afergan, J. Wein, and A. LaMeyer}, ``{Experience with some Principles for
  Building an Internet-Scale Reliable System},'' in \emph{Proceedings of the
  Fifth IEEE International Symposium on Network Computing and Applications
  (NCA'06)}, 2006, p.~3.

\bibitem{Roth2006MDS}
{Ron M. Roth}, \emph{{Introduction to Coding Theory}}, 1st~ed.\hskip 1em plus
  0.5em minus 0.4em\relax Cambridge University Press, 2006, pp. 333--351.

\bibitem{Zakhor2001}
{W. T. Tan and A. Zakhor}, ``{Video Multicast Using Layered FEC and Scalable
  Compression},'' \emph{IEEE Transactions on Circuits and Systems for Video
  Technology}, vol.~11, no.~3, pp. 373--386, 2001.

\bibitem{Dairaine2005}
{L. Dairaine, L. Lancérica, J. Lacan, and J. Fimes}, ``{Content-Access QoS in
  Peer-to-Peer Networks Using a Fast MDS Erasure Code},'' \emph{Elsevier
  Computer Communications}, vol.~28, no.~15, pp. 1778--1790, 2005.

\bibitem{Peng2005}
{X. H. Peng}, ``{Erasure-control Coding for Distributed Networks},'' \emph{IEE
  Proceedings on Communications}, vol. 152, pp. 1075 -- 1080, 2005.

\bibitem{Alon1995}
{N. Alon, J. Edmonds, and M. Luby}, ``{Linear Time Erasure Codes with Nearly
  Optimal Recovery},'' in \emph{IEEE Symposium on Foundations of Computer
  Science, Proc. IEEE Vol. 3}, 1995, pp. 512--519.

\bibitem{Justesen1976}
{J. Justesen }, ``{On the complexity of decoding Reed-Solomon codes},''
  \emph{IEEE transactions on information theory}, vol.~22, no.~2, pp. 237--
  238, 1993.

\bibitem{Luby2001}
{M. G. Luby, M. Mitzenmacher, M. A. Shokrollahi, and D. A. Spielman},
  ``{Efficient Erasure Correcting Codes},'' \emph{IEEE Transactions on
  Information Theory}, vol.~47, no.~2, pp. 569--584, 2001.

\bibitem{Shokrollahi2006}
{A. Shokrollahi}, ``{Raptor Codes},'' \emph{IEEE Transactions on Information
  Theory}, vol.~52, no.~6, pp. 2551--2567, 2006.

\bibitem{Koetter2003}
{R. Koetter and M. Medard }, ``{An algebraic approach to network coding},''
  \emph{IEEE transactions on Networking}, vol.~11, no.~5, pp. 782-- 795, 2003.

\bibitem{Chou2003}
{P. A. Chou, Y. Wu, and K. Jain}, ``{Practical Network Coding },'' in
  \emph{51st Allerton Conference on Communication, Control and Computing},
  2003.

\bibitem{Gkantsidis2005}
{ C. Gkantsidis and P. R. Rodriguez}, ``{Network coding for large scale content
  distribution},'' in \emph{IEEE INFOCOM, Proc. IEEE Vol. 4}, 2005, pp.
  2235--2245.

\bibitem{Yajnik99}
{M. Yajnik, S.B. Moon, J.F. Kurose, and D.F. Towsley }, ``{Measurement and
  Modeling of the Temporal Dependence in Packet Loss},'' in \emph{IEEE INFOCOM
  Proc. IEEE Vol. 1}, 1999, pp. 345--352.

\bibitem{Rossi2003}
{ P. Rossi, G. Romano, F. Palmieri, and G. Iannello }, ``{A Hidden Markov Model
  for Internet Channels},'' in \emph{IEEE International Symposium on Signal
  Processing and Information Technology}, 2003.

\bibitem{Kellerer2002}
{W. Kellerer, E. Steinbach, P. Eisert, and B. Girod}, ``{A Real-Time Internet
  Streaming Media Testbed},'' in \emph{Proc. IEEE International Conference on
  Multimedia and Expo}, 2002, pp. 453-- 456.

\bibitem{Henocq2000}
{X. Henocq and C. Guillemot}, ``{Source Adaptive Error Control for Real-time
  Video over the Internet},'' \emph{Numéro spécial image et vidéo. Hermès.
  Réseaux et système réparti. Calculateurs Parallèles}, vol.~12, no. 3-4,
  2000.

\bibitem{Leannec19992}
{F. L. Leannec and C. Guillemot}, ``{Error Resilient Video Transmission over
  the Internet},'' in \emph{Proc. Visual Communication and Image Processing},
  1999.

\bibitem{Salamatian2001}
{K. Salamatian and Vaton}, ``{Hidden Markov Modeling for Network Communication
  Channels},'' in \emph{Proc. ACM SIGMETRICS}, 2001, pp. 92 -- 101.

\bibitem{Roth2006RSdec}
{Ron M. Roth}, \emph{{Introduction to Coding Theory}}, 1st~ed.\hskip 1em plus
  0.5em minus 0.4em\relax Cambridge University Press, 2006, pp. 183--204.

\bibitem{Padhye2000}
{J. Padhye, V. Firoiu, D.F Towsley, and J.F. Kurose}, ``{Modeling TCP Reno
  performance: a simple model and its empirical validation},'' \emph{IEEE/ACM
  Transactions on Networking}, vol.~8, no.~2, pp. 133 -- 145, 2000.

\bibitem{Dembo1998}
{Amir Dembo and Ofer Zeitouni}, \emph{{Large Deviations Techniques and
  Applications}}, 2nd~ed.\hskip 1em plus 0.5em minus 0.4em\relax New York:
  Springer, 1998, pp. 11--43.

\bibitem{Kelley1975}
{J. L. Kelley}, \emph{{General Topology}}.\hskip 1em plus 0.5em minus
  0.4em\relax Springer, 1975, pp. 40--43.

\bibitem{Papadimitriou1994}
{C. H. Papadimitriou}, \emph{{Computational Complexity}}, 1st~ed.\hskip 1em
  plus 0.5em minus 0.4em\relax New York: Addison Wesley, 1994.

\bibitem{Cormen2001}
{T. H. Cormen, C. E. Leiserson, R. L. Rivest, and C. Stein},
  \emph{{Introduction to Algorithms}}, 2nd~ed.\hskip 1em plus 0.5em minus
  0.4em\relax MIT Press, 2001, pp. 347--349.

\bibitem{Rudin1976}
{W. Rudin}, \emph{{Principles of Mathematical Analysis}}, 3rd~ed.\hskip 1em
  plus 0.5em minus 0.4em\relax McGraw-Hill, 1976.

\bibitem{Boyd2004}
{S. Boyd and L. Vandenberghe}, \emph{{Convex Optimization}}, 1st~ed.\hskip 1em
  plus 0.5em minus 0.4em\relax Cambridge, UK: Cambridge University Press, 2004,
  pp. 243--245.

\bibitem{Cover1991Sanov}
{T. Cover and J. Thomas}, \emph{{Elements of Information Theory}},
  1st~ed.\hskip 1em plus 0.5em minus 0.4em\relax New York: Wiley, 1991, pp.
  291--294.

\bibitem{Cover1991Distance}
------, \emph{{Elements of Information Theory}}, 1st~ed.\hskip 1em plus 0.5em
  minus 0.4em\relax New York: Wiley, 1991, pp. 30--31.

\end{thebibliography}

\end{document}